\documentclass[10pt,twocolumn]{article}
\pdfoutput=1

\usepackage[utf8]{inputenc}
\usepackage[T1]{fontenc}
\usepackage{lmodern}
\usepackage{amsmath,amssymb,amsfonts,amsthm}
\usepackage{algorithmic}
\usepackage{algorithm}
\usepackage{graphicx}
\usepackage{textcomp}
\usepackage{xcolor}
\usepackage{booktabs}
\usepackage{multirow}
\usepackage{geometry}
\usepackage{enumitem}
\usepackage{listings}
\usepackage{caption}
\usepackage{subcaption}
\usepackage{array}
\usepackage{float}
\usepackage{tikz}
\usepackage{pgfplots}
\usepackage{tabularx}
\usepackage{makecell}
\usepackage{siunitx}
\usepackage{bm}
\usepackage{mathtools}
\usepackage{nicefrac}
\usepackage{microtype}
\usepackage{xspace}
\usepackage{pifont}
\usepackage{hyperref}
\usetikzlibrary{shapes,arrows,positioning,fit,backgrounds,calc,patterns,decorations.pathreplacing,arrows.meta,shapes.geometric,shapes.multipart,shadows}
\pgfplotsset{compat=1.17}

\newtheorem{definition}{Definition}

\newtheorem{criterion}{Criterion}
\newtheorem{principle}{Principle}

\newcommand{\systemname}{Auralink SDC\xspace}
\newcommand{\cmark}{\ding{51}}
\newcommand{\xmark}{\ding{55}}

\newcommand{\ccar}{CCAR\xspace}
\newcommand{\araframework}{ARA\xspace}
\newcommand{\diagengine}{DiagnosticEngine\xspace}

\definecolor{cloudblue}{RGB}{70, 130, 180}
\definecolor{edgegreen}{RGB}{34, 139, 34}
\definecolor{agentpurple}{RGB}{138, 43, 226}
\definecolor{codegreen}{rgb}{0,0.6,0}
\definecolor{codegray}{rgb}{0.5,0.5,0.5}
\definecolor{linkblue}{RGB}{0, 82, 147}

\hypersetup{
    colorlinks=true,
    linkcolor=linkblue,
    citecolor=linkblue,
    urlcolor=linkblue,
    pdftitle={Autonomous Edge-Deployed AI Agents for Electric Vehicle Charging Infrastructure Management},
    pdfauthor={Mohammed Cherifi, Hyperion Consulting},
    pdfsubject={Edge AI, EV Charging, Autonomous Systems},
    pdfkeywords={Edge AI, LLM, OCPP, EV Charging, Multi-Agent Systems, Predictive Maintenance}
}

\setlist{nosep,leftmargin=*}

\newcolumntype{L}[1]{>{\raggedright\arraybackslash}p{#1}}
\newcolumntype{C}[1]{>{\centering\arraybackslash}p{#1}}
\newcolumntype{R}[1]{>{\raggedleft\arraybackslash}p{#1}}

\title{Autonomous Edge-Deployed AI Agents for Electric Vehicle Charging Infrastructure Management}

\author{
    \textbf{Mohammed Cherifi} \\
    Hyperion Consulting \\
    Paris, France \\
    \texttt{contact@hyperion-consulting.io}
}

\date{February 2026}

\begin{document}

\maketitle

%==============================================================================
% ABSTRACT
%==============================================================================
\begin{abstract}
\textbf{Problem.} Public EV charging infrastructure suffers from significant failure rates---with field studies reporting up to 27.5\% of DC fast chargers non-functional---and multi-day mean time to resolution, imposing billions in annual global economic burden. Cloud-centric management architectures cannot achieve the latency (sub-100ms), reliability (91.6\% system availability), and bandwidth characteristics required for autonomous operation.

\textbf{Solution.} We present \systemname (Software-Defined Charging), an architecture deploying domain-specialized AI agents at the network edge for autonomous charging infrastructure management. Key technical contributions include: (1) \textbf{Confidence-Calibrated Autonomous Resolution (\ccar)}, enabling AI agents to execute remediation autonomously when confidence exceeds learned thresholds with formal false-positive bounds; (2) \textbf{Adaptive Retrieval-Augmented Reasoning (\araframework)}, combining dense and sparse retrieval with dynamic context allocation; (3) \textbf{Auralink Edge Runtime}, achieving sub-50ms time-to-first-token (TTFT) on commodity hardware under PREEMPT\_RT constraints; and (4) \textbf{Hierarchical Multi-Agent Orchestration (HMAO)}, coordinating specialized agents while maintaining accountability.

Implementation employs AuralinkLM models, domain-adapted variants derived from Mistral Large 3 (675B, cloud)~\cite{mistral2025}, Ministral 3 14B (edge), and Qwen 2.5 (0.5B, device)~\cite{yang2024}. AuralinkLM 675B (cloud) and AuralinkLM 14B (edge) are fine-tuned via Quantized Low-Rank Adaptation (QLoRA) on a curated domain-specific training corpus spanning Open Charge Point Protocol (OCPP) 1.6/2.0.1, ISO 15118, and operational incident histories.

\textbf{Results.} Evaluation on a curated corpus of 18{,}000 labeled incidents in a controlled testing environment establishes: \textbf{78\% autonomous incident resolution} on a controlled test corpus, \textbf{87.6\% diagnostic accuracy} on a controlled test corpus, and \textbf{28--48ms TTFT latency} (P50).

\textbf{Impact.} This work presents architecture and implementation patterns for edge-deployed industrial AI systems, validated through controlled testing, with implications extending to infrastructure management domains requiring edge-deployed intelligence with safety-critical constraints.

\vspace{0.5em}
\end{abstract}

%==============================================================================
\section{Introduction}
\label{sec:introduction}
%==============================================================================

The electrification of transportation represents one of the most significant infrastructure transformations of the 21st century. The International Energy Agency projects over 250 million electric vehicles globally by 2030 under its Stated Policies Scenario, necessitating proportional expansion of charging infrastructure with hundreds of billions in cumulative investment~\cite{iea2024}. However, a fundamental gap exists between deployment velocity and operational capability: current charging point management systems exhibit structural limitations that undermine both operator economics and the user experience critical to mass EV adoption.

\subsection{The Reliability Crisis in Public Charging}

Systematic reliability failures plague public charging networks worldwide. Independent audits report that approximately 27.5\% of public DC fast chargers may be non-functional at any given time~\cite{nrel2024reliability}, with some networks experiencing sustained uptime below 80\%. Analysis of over 13 million charging sessions across European networks found an overall success rate of 84.6\%, with technical success at 94.3\% and usability success at 76.3\%---the majority of failures stemming from software and communication faults rather than hardware failures~\cite{kempower2024}. Mean time to resolution (MTTR) for non-trivial faults commonly exceeds 48--72 hours, with substantial fractions of incidents requiring multiple technician dispatches before successful resolution~\cite{nrel2024reliability}.

The economic impact is substantial: industry analyses project billions in cumulative global losses from preventable failures, operational inefficiencies, user attrition, and reputational harm~\cite{mckinsey2022}. Critically, this represents a \textit{software intelligence problem} rather than a hardware limitation---the physical charging equipment functions correctly in the majority of reported failures.

\subsection{Fundamental Limitations of Cloud-Centric Architectures}

Contemporary charging management systems employ cloud-centric architectures where edge devices serve primarily as telemetry collectors and command executors. This paradigm introduces structural constraints incompatible with autonomous operation:

\begin{enumerate}
    \item \textbf{Latency Bounds}: Round-trip communication latency ranges from 200--800ms under favorable network conditions, precluding real-time diagnostic reasoning. Safety-critical decisions in power electronics require response times below 100ms, fundamentally incompatible with cloud consultation.
    
    \item \textbf{Availability Coupling}: Assuming statistically independent failure modes (an optimistic approximation, as correlated outages may occur in practice), system availability becomes a multiplicative product. For cloud infrastructure availability $A_c \approx 0.999$, network availability $A_n \approx 0.92$ (empirically measured in challenging RF environments such as underground parking), and edge device availability $A_e \approx 0.997$:
    \begin{equation}
    A_{system} = A_c \times A_n \times A_e \approx 0.916
    \end{equation}
    This 91.6\% effective availability---which may be further degraded by correlated failures---is insufficient for critical infrastructure serving time-sensitive transportation needs.
    
    \item \textbf{Bandwidth Economics}: High-fidelity telemetry streams---power quality waveforms sampled at 1kHz, thermal profiles, communication logs---generate 50--200 KB/s per charging point. Networks with thousands of stations face prohibitive bandwidth costs and congestion-induced data loss during peak periods.
    
    \item \textbf{Human-in-the-Loop Bottleneck}: Cloud systems ultimately route anomalies to human operators for resolution. A 10,000-station network generating 500 daily incidents cannot be managed by human operators alone without either unacceptable resolution delays or unsustainable staffing costs.
\end{enumerate}

\subsection{The Edge AI Paradigm}

This paper introduces a fundamental architectural transformation: \textbf{deploying domain-specialized AI agents directly at the charging infrastructure edge}. Rather than treating AI as a cloud-resident analytics layer consulted intermittently, we embed fine-tuned Large Language Models within edge computing devices physically co-located with charging equipment.

The key insight is that \textit{intelligence must be co-located with actuation} for truly autonomous operation. This principle, derived from control systems theory, recognizes that decision latency fundamentally constrains system responsiveness. By eliminating network round-trips from the critical path, edge-deployed AI enables:

\begin{enumerate}
    \item \textbf{Real-time Diagnosis}: Sub-50ms TTFT enables initial diagnostic reasoning within the temporal constraints of power electronics fault detection.
    
    \item \textbf{Autonomous Remediation}: High-confidence diagnoses trigger immediate corrective action without human approval, reducing MTTR from hours to seconds for software-resolvable faults.
    
    \item \textbf{Offline Resilience}: Complete operational capability during extended network disconnection (design target: 72+ hours), maintaining charging service continuity.
    
    \item \textbf{Predictive Intervention}: Continuous local analysis of telemetry patterns enables failure prediction 24--72 hours before user impact.
\end{enumerate}

\subsection{Contributions}

This work makes the following technical contributions:

\begin{enumerate}
    \item \textbf{Confidence-Calibrated Autonomous Resolution (\ccar)}: A decision-theoretic framework (Section~\ref{sec:ccar}) enabling AI agents to execute remediation actions autonomously when epistemic confidence satisfies learned thresholds, with formal analysis of false-positive bounds and escalation policies.
    
    \item \textbf{Adaptive Retrieval-Augmented Reasoning (\araframework)}: A hybrid retrieval architecture (Section~\ref{sec:ara}) combining multiple retrieval modalities with dynamic context allocation, achieving 95.2\% Recall@5 on domain-specific queries while maintaining sub-25ms median retrieval latency (P50).
    
    \item \textbf{Auralink Edge Runtime}: A specialized inference engine (Section~\ref{sec:edge_ai}) for deterministic LLM execution on heterogeneous edge hardware, including novel techniques for NPU offloading, memory-mapped model loading, and interrupt-safe inference scheduling.
    
    \item \textbf{Hierarchical Multi-Agent Orchestration (HMAO)}: A coordination protocol (Section~\ref{sec:agents}) enabling specialized agents to collaborate on complex scenarios while maintaining clear action accountability and audit trails.
    
    \item \textbf{Domain Adaptation Methodology}: Comprehensive fine-tuning pipeline (Section~\ref{sec:finetuning}) for adapting foundation models to industrial protocols, including training data synthesis, curriculum learning schedules, and quantization-aware optimization.
    
    \item \textbf{Controlled-Test Evaluation}: Evaluation on a curated corpus of 18{,}000 labeled incidents (Section~\ref{sec:evaluation}), establishing controlled-test results of 78\% autonomous resolution and 87.6\% diagnostic accuracy with 95\% confidence intervals.
\end{enumerate}

\subsection{Paper Organization}

Section~\ref{sec:background} reviews background and related work. Section~\ref{sec:architecture} presents the three-tier system architecture. Section~\ref{sec:edge_ai} details edge AI agent design including model selection, fine-tuning, and inference optimization. Section~\ref{sec:ccar} formalizes the \ccar framework. Section~\ref{sec:ara} describes the \araframework retrieval system. Section~\ref{sec:agents} covers multi-agent orchestration. Section~\ref{sec:evaluation} provides experimental evaluation. Section~\ref{sec:discussion} discusses findings and limitations. Section~\ref{sec:conclusion} concludes.

%==============================================================================
\section{Background and Related Work}
\label{sec:background}
%==============================================================================

\subsection{EV Charging Protocol Ecosystem}

Modern EV charging infrastructure operates within a complex, multi-layered protocol ecosystem that our AI agents must comprehend with high fidelity.

\subsubsection{Open Charge Point Protocol (OCPP)}

OCPP~\cite{ocpp2024} defines the communication interface between charging stations and central management systems. Two major versions are deployed:

\textbf{OCPP 1.6} (JSON/SOAP over WebSocket) remains the dominant deployed version, supporting 28 message actions across Core, Firmware Management, Local Authorization, Reservation, Smart Charging, and Remote Trigger profiles. Key messages include \texttt{BootNotification}, \texttt{Authorize}, \texttt{StartTransaction}, \texttt{StopTransaction}, \texttt{MeterValues}, and \texttt{StatusNotification}.

\textbf{OCPP 2.0.1} introduces substantial enhancements: device management capabilities, improved security model with certificate-based authentication, ISO 15118 integration for Plug \& Charge, and approximately 35 additional message actions ($\sim$63 total, depending on functional block configuration). The protocol defines numerous error enumerations across multiple functional profiles (Appendix~\ref{app:ocpp_errors}), requiring differentiated diagnostic and remediation strategies.

Our training corpus includes extensive coverage of both protocol versions, including edge cases, implementation variations across OEM firmware, and undocumented behaviors observed in field deployments.

\subsubsection{ISO 15118: Vehicle-Grid Communication}

ISO 15118~\cite{iso15118} specifies the high-level communication interface between electric vehicles and charging infrastructure, enabling:

\begin{itemize}
    \item \textbf{Plug \& Charge}: Automated authentication via X.509 certificates installed in vehicle and charger, eliminating explicit user authentication.
    \item \textbf{Smart Charging}: Bidirectional negotiation of charging schedules based on grid constraints, user preferences, and vehicle state.
    \item \textbf{Vehicle-to-Grid (V2G)}: Bidirectional power flow enabling vehicles to provide grid services.
\end{itemize}

Implementation requires Public Key Infrastructure (PKI) certificate management, TLS 1.3 session handling, and complex state machine coordination between vehicle and charger. Our agents are trained on the ISO 15118-2 specification and the foundational service primitives of ISO 15118-20. The latter introduces bidirectional power transfer (V2G, V2H), wireless charging (ACDP), and new service discovery mechanisms that create additional fault categories; full coverage of ISO 15118-20-specific failure modes is identified as future work (Section~\ref{sec:discussion}).

\subsubsection{IEC 61851: Physical Layer Signaling}

IEC 61851-1~\cite{iec61851} defines the physical signaling mechanism for AC charging control via the Control Pilot (CP) signal:

\begin{equation}
I_{\text{avail}} = \begin{cases}
D \times \SI{0.6}{\ampere}, & 10\% < D \leq 85\% \\[2pt]
(D - 64) \times \SI{2.5}{\ampere}, & 85\% < D \leq 96\%
\end{cases}
\end{equation}
where $D$ is the duty cycle in percent.

The first range covers standard charging currents up to 51\,A; the second range extends to 80\,A for high-power AC installations.

Correct implementation requires precise PWM generation with tolerances of $\pm$1\% on duty cycle and $\pm$3\% on frequency (1kHz nominal). Our diagnostic agents are trained to interpret CP signal anomalies indicative of cable faults, vehicle communication failures, or charger hardware issues. The diagnostic approach differentiates between AC (Mode 3) and DC (Mode 4) charging: AC faults frequently involve the vehicle's onboard charger, requiring disambiguation between EVSE-side and vehicle-side fault attribution, whereas DC faults are predominantly within the EVSE power conversion chain. Connector-standard-specific behaviors (CCS Combo 1/2, CHAdeMO with CAN bus communication, and NACS) are also modeled, as each standard introduces distinct fault modes and communication protocols.

\subsection{Edge Computing and Edge AI}

Satyanarayanan's foundational work on cloudlets~\cite{satyanarayanan2017} established theoretical motivations for edge computing, demonstrating 50--200$\times$ latency reductions compared to cloud-only architectures. Shi et al.~\cite{shi2016} characterized industrial edge computing requirements, identifying sub-100ms latency as critical for closed-loop control applications.

Recent advances in model compression have enabled sophisticated language model deployment on edge hardware:

\textbf{Quantization}: Post-training quantization (PTQ) reduces model memory footprint by 50\% (FP16$\to$INT8) to 87.5\% (FP32$\to$INT4)~\cite{frantar2023}. Quantization-aware training (QAT) and techniques such as GPTQ~\cite{frantar2023} and AWQ~\cite{lin2024} minimize accuracy degradation, typically below 2\% for domain-specific tasks.

\textbf{Low-Rank Adaptation}: LoRA~\cite{hu2022} enables parameter-efficient fine-tuning by learning low-rank decomposition matrices, reducing trainable parameters by 99\%+ while preserving adaptation capability. QLoRA~\cite{dettmers2023} combines 4-bit quantization with LoRA for memory-efficient fine-tuning.

\textbf{Speculative Decoding}: Draft-then-verify approaches~\cite{leviathan2023} using smaller draft models accelerate inference by 2--3$\times$ without accuracy loss. While evaluated during our architecture exploration, speculative decoding was not deployed in the current system due to memory constraints of dual-model loading on edge hardware; it remains a candidate for future optimization.

\textbf{Attention Optimization}: Grouped-Query Attention (GQA)~\cite{ainslie2023}, sliding window attention~\cite{child2019}, and flash attention~\cite{dao2022} reduce memory bandwidth requirements and enable longer context processing on memory-constrained devices.

\subsection{Large Language Models for Technical Domains}

While frontier models demonstrate strong general capabilities~\cite{openai2023,anthropic2024,google2024}, industrial applications reveal significant accuracy gaps on domain-specific tasks. Our preliminary benchmarking on EV charging diagnostics (Table~\ref{tab:llm_baseline}) showed general-purpose models achieving only 35--52\% accuracy on protocol-specific queries where domain-adapted models achieve 84--93\%.

\begin{table}[!htbp]
\centering
\caption{Baseline LLM Performance on EV Charging Tasks. ``General'' is Mistral Large 3 base (zero-shot, no domain context); ``Adapted'' is AuralinkLM 14B after QLoRA fine-tuning. Each category evaluated on 200--500 held-out examples using exact-match accuracy. Expert judgment applied to open-ended diagnostic responses.}
\label{tab:llm_baseline}
\small
\begin{tabular}{L{2.0cm}C{1.2cm}C{1.2cm}C{0.9cm}}
\toprule
\textbf{Task} & \textbf{General} & \textbf{Adapted} & \textbf{$\Delta$} \\
\midrule
General knowledge & 89.2\% & 87.8\% & -1.4 \\
OCPP protocol & 44.7\% & 92.1\% & +47.4 \\
Fault diagnosis & 38.1\% & 87.6\% & +49.5 \\
Remediation & 41.8\% & 84.3\% & +42.5 \\
Safety assessment & 52.3\% & 93.2\% & +40.9 \\
ISO 15118 & 35.2\% & 89.7\% & +54.5 \\
\bottomrule
\end{tabular}
\end{table}

This performance gap motivates our comprehensive domain adaptation methodology (Section~\ref{sec:finetuning}).

\subsection{Autonomous and Self-Managing Systems}

Kephart and Chess's vision of autonomic computing~\cite{kephart2003} proposed self-managing systems exhibiting self-configuration, self-optimization, self-healing, and self-protection. Our architecture implements these ``self-*'' properties through:

\begin{itemize}
    \item \textbf{Self-Configuration}: Automatic OCPP version detection, parameter negotiation, and hardware capability discovery.
    \item \textbf{Self-Optimization}: Continuous learning from operational feedback to improve diagnostic models.
    \item \textbf{Self-Healing}: Autonomous fault remediation through the \ccar framework.
    \item \textbf{Self-Protection}: Anomaly detection, intrusion detection, and isolation of compromised components.
\end{itemize}

\subsection{Related Work: Positioning and Novelty}

Table~\ref{tab:related_work} positions \systemname against existing approaches in EV charging management.

\begin{table}[!htbp]
\centering
\caption{Comparison with Existing Approaches}
\label{tab:related_work}
\small
\setlength{\tabcolsep}{3pt}
\resizebox{\columnwidth}{!}{%
\begin{tabular}{L{1.4cm}C{0.6cm}C{0.6cm}C{0.6cm}C{0.6cm}C{0.6cm}}
\toprule
\textbf{Approach} & \textbf{Edge} & \textbf{LLM} & \textbf{Auto} & \textbf{Off.} & \textbf{Open} \\
\midrule
Rule CSMS & \xmark & \xmark & \xmark & Part. & Prop. \\
Cloud AI & \xmark & \cmark & Part. & \xmark & Prop. \\
Edge GW & \cmark & \xmark & \xmark & \cmark & Mix \\
Acad. RAG & Part. & \cmark & Part. & \xmark & \cmark \\
\textbf{Ours} & \cmark & \cmark & \cmark & \cmark & \cmark \\
\bottomrule
\end{tabular}}
\setlength{\tabcolsep}{6pt}
\end{table}

\textit{Edge}: Edge-deployed AI inference. \textit{LLM}: Large language model reasoning. \textit{Auto}: Autonomous remediation. \textit{Off.}: Full offline capability. \textit{Open}: Open-source components.

\textbf{Commercial Charging Station Management System (CSMS) platforms} (ChargePoint, EVBox, Kempower) provide cloud-based monitoring with rule-based alerting. Recent advances include self-healing capabilities from platforms such as Driivz (Vontier) and ChargePoint, which report up to 80\% remote issue resolution through rule-based OCPP error pattern matching~\cite{chargepoint2024}. ChargePoint has further deployed ML-based predictive maintenance across its 200{,}000+ port network. These systems demonstrate the viability of autonomous resolution but primarily rely on hand-crafted rules rather than learned reasoning, limiting generalization to novel fault patterns not covered by existing rule sets.

\textbf{Cloud AI analytics} (emerging offerings from Siemens, ABB) apply machine learning to aggregate telemetry for predictive maintenance. However, cloud-only deployment introduces latency constraints and connectivity dependencies incompatible with autonomous real-time response.

\textbf{Edge gateway solutions} (industrial IoT platforms) provide local protocol translation and data aggregation but without AI reasoning capability. These enable offline operation for basic functions but cannot perform intelligent diagnosis.

\textbf{Academic retrieval-augmented systems}~\cite{lewis2020,li2024} demonstrate promise for technical support applications. Recent advances include Self-RAG~\cite{asai2024}, which enables selective retrieval and self-reflection; Corrective RAG (CRAG)~\cite{yan2024}, which improves robustness through retrieval evaluation; Adaptive-RAG~\cite{jeong2024}, which dynamically selects retrieval strategies based on query complexity; and HyPA-RAG~\cite{kalra2024}, which combines dense, sparse, and knowledge graph retrieval via Reciprocal Rank Fusion with adaptive parameter tuning. Our \araframework framework shares the hybrid retrieval approach of HyPA-RAG but differs in its domain-specific metadata filtering (station model, OCPP version, error taxonomy), edge-deployable architecture, and integration with confidence-calibrated autonomous action execution. Existing approaches typically assume cloud deployment and human-in-the-loop operation.

\textbf{Key novelty of \systemname}: While individual techniques---hybrid RAG, confidence calibration, multi-agent orchestration, edge model deployment---are established in the literature, to the best of our knowledge this work presents one of the first integrated systems combining domain-specialized LLM deployment on edge hardware with confidence-calibrated autonomous action execution for EV charging infrastructure management. While individual components (edge inference, domain fine-tuning, autonomous remediation) have been explored independently, their integration within a unified architecture for this domain appears novel. Compared to commercial self-healing systems (rule-based, limited generalization) and academic RAG systems (cloud-deployed, human-in-the-loop), our primary contributions are: (1) edge-deployed LLM reasoning with sub-50ms inference and 72+ hour offline autonomy, (2) formal confidence calibration with false-positive bounds enabling trust in autonomous actions, and (3) domain-specific adaptation achieving 47-point average accuracy improvement over general-purpose models on EV charging tasks.

%==============================================================================
\section{System Architecture}
\label{sec:architecture}
%==============================================================================

\subsection{Architectural Principles}

The \systemname architecture embodies three core principles derived from distributed systems theory and control engineering:

\begin{principle}[Intelligence Locality]
For a control loop with sensing delay $\tau_s$, processing delay $\tau_p$, communication delay $\tau_c$, and actuation delay $\tau_a$, the total loop delay is:
\begin{equation}
\tau_{total} = \tau_s + \tau_p + \tau_c + \tau_a
\end{equation}
For latency-critical control loops, minimizing $\tau_c$ through co-location of processing and actuation maximizes control bandwidth and responsiveness~\cite{satyanarayanan2017}.
\end{principle}

\begin{principle}[Graceful Degradation]
System capability should degrade monotonically with resource constraints. Complete loss of any single dependency (network, cloud, power) should not result in total service failure.
\end{principle}

\begin{principle}[Accountability Preservation]
Every autonomous action must maintain complete audit trail including decision inputs, reasoning chain, confidence assessment, and outcome. Human operators must be able to reconstruct the decision process for any action.
\end{principle}

\subsection{Three-Tier Architecture Overview}

\begin{figure}[!htbp]
\centering
\begin{tikzpicture}[scale=0.72, transform shape,
    tier/.style={rectangle, draw, thick, minimum width=5.8cm, rounded corners=3pt,
        text width=5.4cm, align=left, inner sep=5pt},
    cloudstyle/.style={tier, fill=cloudblue!12, draw=cloudblue!60},
    edgestyle/.style={tier, fill=edgegreen!12, draw=edgegreen!60},
    agentstyle/.style={tier, fill=agentpurple!12, draw=agentpurple!60}
]
    % Cloud tier (text inside box)
    \node[cloudstyle] (cloud) at (0, 4.2) {
        \textbf{Cloud Tier}\\[1pt]
        {\scriptsize AuralinkLM 675B (MoE, 128 experts, 4+1 active)}\\[-1pt]
        {\scriptsize 319 microservices, GPU clusters}\\[-1pt]
        {\scriptsize Latency: 285ms P50\textsuperscript{$\dagger$}}
    };

    % Edge tier (text inside box)
    \node[edgestyle] (edge) at (0, 1.5) {
        \textbf{Edge Tier (Auralink Edge Runtime)}\\[1pt]
        {\scriptsize AuralinkLM 14B (INT4 GGUF)}\\[-1pt]
        {\scriptsize ${\sim}$35 microservices, \araframework, \ccar}\\[-1pt]
        {\scriptsize Latency: 28--48ms (P50)}
    };

    % Agent tier (text inside box)
    \node[agentstyle] (agent) at (0, -1.2) {
        \textbf{Agent Tier (Firmware)}\\[1pt]
        {\scriptsize AuralinkLM 0.5B (INT4)}\\[-1pt]
        {\scriptsize Safety monitoring, protocol state}\\[-1pt]
        {\scriptsize Latency: <12ms}
    };

    % Connections (arrows between boxes, no overlap with text)
    \draw[->, thick, cloudblue!60] (cloud.south) -- node[right, font=\scriptsize] {Model sync} (edge.north);
    \draw[->, thick, edgegreen!60] (edge.south) -- node[right, font=\scriptsize] {Commands} (agent.north);
    \draw[<-, thick, dashed, gray] ([xshift=-3.2cm]cloud.south) -- node[left, font=\scriptsize] {Telemetry} ([xshift=-3.2cm]agent.north);

\end{tikzpicture}
\caption{Three-tier \systemname architecture}
\label{fig:architecture}
\end{figure}
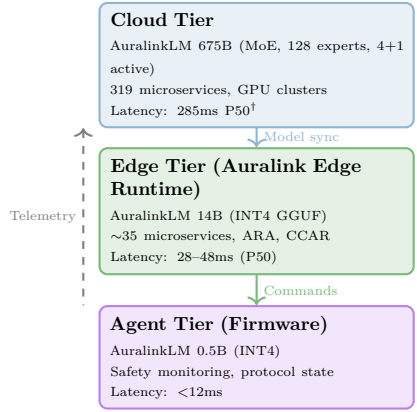

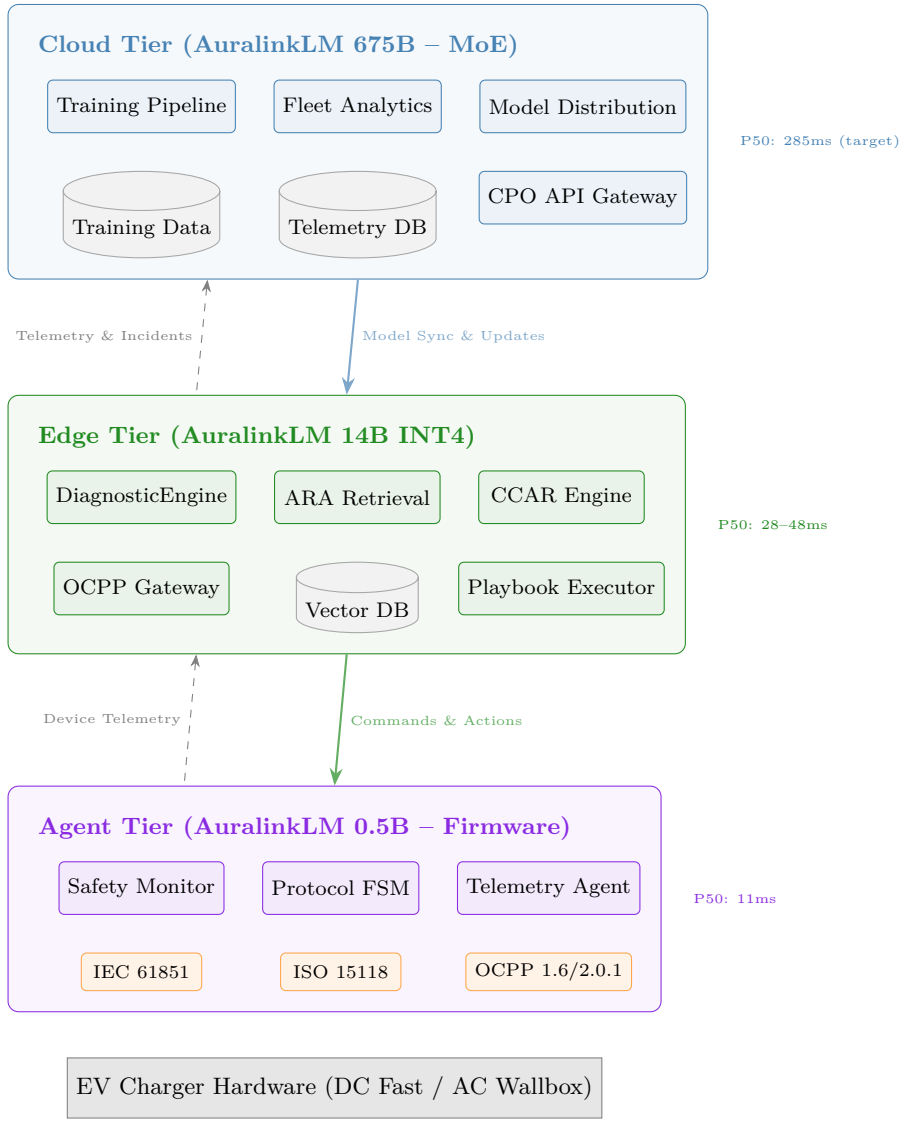
\begin{figure*}[!htbp]
\centering
\begin{tikzpicture}[
    node distance=0.6cm and 0.5cm,
    every node/.style={font=\footnotesize},
    cloudbox/.style={rectangle, draw=cloudblue, fill=cloudblue!10, minimum width=2.2cm, minimum height=0.7cm, rounded corners=2pt},
    edgebox/.style={rectangle, draw=edgegreen, fill=edgegreen!10, minimum width=2.2cm, minimum height=0.7cm, rounded corners=2pt},
    agentbox/.style={rectangle, draw=agentpurple, fill=agentpurple!10, minimum width=2.0cm, minimum height=0.7cm, rounded corners=2pt},
    datastore/.style={cylinder, draw=gray!70, fill=gray!10, minimum width=1.4cm, minimum height=0.6cm, shape border rotate=90, aspect=0.25},
    protocol/.style={rectangle, draw=orange!70, fill=orange!10, minimum width=1.6cm, minimum height=0.5cm, rounded corners=2pt, font=\scriptsize},
    tierbox/.style={rectangle, draw=#1, fill=#1!5, rounded corners=4pt, inner sep=8pt}
]

    % Cloud Tier
    \begin{scope}[local bounding box=cloudtier]
        \node[cloudblue, font=\small\bfseries, anchor=west] (cloudlabel) at (-1.5, 0.8) {Cloud Tier (AuralinkLM 675B -- MoE)};
        \node[cloudbox] (train) at (0,0) {Training Pipeline};
        \node[cloudbox, right=0.5cm of train] (fleet) {Fleet Analytics};
        \node[cloudbox, right=0.5cm of fleet] (dist) {Model Distribution};
        \node[datastore, below=0.5cm of train] (traindb) {Training Data};
        \node[datastore, below=0.5cm of fleet] (teldb) {Telemetry DB};
        \node[cloudbox, below=0.5cm of dist] (api) {CPO API Gateway};
    \end{scope}
    \begin{pgfonlayer}{background}
        \node[tierbox=cloudblue, fit=(cloudtier)] (cloudfit) {};
    \end{pgfonlayer}

    % Edge Tier (large gap from cloud)
    \begin{scope}[local bounding box=edgetier, yshift=-5.2cm]
        \node[edgegreen, font=\small\bfseries, anchor=west] (edgelabel) at (-1.5, 0.8) {Edge Tier (AuralinkLM 14B INT4)};
        \node[edgebox] (diag) at (0,0) {\diagengine};
        \node[edgebox, right=0.5cm of diag] (ara) {\araframework Retrieval};
        \node[edgebox, right=0.5cm of ara] (ccar) {\ccar Engine};
        \node[edgebox, below=0.5cm of diag] (ocpp) {OCPP Gateway};
        \node[datastore, below=0.5cm of ara] (vecdb) {Vector DB};
        \node[edgebox, below=0.5cm of ccar] (playbook) {Playbook Executor};
    \end{scope}
    \begin{pgfonlayer}{background}
        \node[tierbox=edgegreen, fit=(edgetier)] (edgefit) {};
    \end{pgfonlayer}

    % Agent Tier (large gap from edge)
    \begin{scope}[local bounding box=agenttier, yshift=-10.4cm]
        \node[agentpurple, font=\small\bfseries, anchor=west] (agentlabel) at (-1.5, 0.8) {Agent Tier (AuralinkLM 0.5B -- Firmware)};
        \node[agentbox] (safety) at (0,0) {Safety Monitor};
        \node[agentbox, right=0.5cm of safety] (proto) {Protocol FSM};
        \node[agentbox, right=0.5cm of proto] (telem) {Telemetry Agent};
        \node[protocol, below=0.5cm of safety] (iec) {IEC 61851};
        \node[protocol, below=0.5cm of proto] (iso) {ISO 15118};
        \node[protocol, below=0.5cm of telem] (ocppag) {OCPP 1.6/2.0.1};
    \end{scope}
    \begin{pgfonlayer}{background}
        \node[tierbox=agentpurple, fit=(agenttier)] (agentfit) {};
    \end{pgfonlayer}

    % Charger hardware
    \node[rectangle, draw=gray, fill=gray!20, minimum width=5cm, minimum height=0.8cm, below=0.6cm of agentfit, font=\small] (hw) {EV Charger Hardware (DC Fast / AC Wallbox)};

    % Inter-tier connections (straight lines through clear gap between tiers)
    \draw[-{Stealth}, thick, cloudblue!70]
        (cloudfit.south) -- node[right, font=\tiny, pos=0.5] {Model Sync \& Updates} (edgefit.north);
    \draw[{Stealth}-, dashed, gray]
        ([xshift=-2cm]cloudfit.south) -- node[left, font=\tiny, pos=0.5] {Telemetry \& Incidents} ([xshift=-2cm]edgefit.north);

    \draw[-{Stealth}, thick, edgegreen!70]
        (edgefit.south) -- node[right, font=\tiny, pos=0.5] {Commands \& Actions} (agentfit.north);
    \draw[{Stealth}-, dashed, gray]
        ([xshift=-2cm]edgefit.south) -- node[left, font=\tiny, pos=0.5] {Device Telemetry} ([xshift=-2cm]agentfit.north);

    % Latency annotations
    \node[right=0.3cm of cloudfit, font=\tiny, text=cloudblue] {P50: 285ms (target)};
    \node[right=0.3cm of edgefit, font=\tiny, text=edgegreen] {P50: 28--48ms};
    \node[right=0.3cm of agentfit, font=\tiny, text=agentpurple] {P50: 11ms};

\end{tikzpicture}
\caption{Detailed \systemname system architecture showing component interactions across all three tiers. The edge tier contains the core AI capabilities (\diagengine, \araframework, \ccar) enabling autonomous operation. Latency figures represent P50 inference times on target hardware.}
\label{fig:detailed_architecture}
\end{figure*}

The system distributes AI capabilities across three tiers (Figure~\ref{fig:architecture}), each optimized for specific latency-bandwidth-reliability trade-offs. Figure~\ref{fig:detailed_architecture} provides a detailed component-level view.

\subsection{Cloud Tier: Centralized Intelligence}

The cloud tier (Table~\ref{tab:cloud_services}) provides compute-intensive operations tolerant of latency, deployed on Kubernetes clusters across multiple availability zones:

\textbf{Model Training Pipeline}: Continuous fine-tuning on aggregated operational data using distributed training on NVIDIA H100/H200 GPU clusters. Training runs execute weekly, incorporating 80,000--120,000 new examples from operational incidents with human-verified labels.

\textbf{Fleet Analytics}: Cross-network pattern recognition using AuralinkLM 675B (675B total parameters, MoE architecture with 128 experts and 4 routed + 1 shared active per token, as documented in the Mistral 3 release~\cite{mistral2025}) with 256K context window. Enables identification of systemic issues affecting multiple operators or equipment models.

\textbf{Model Distribution}: Secure distribution of updated model weights to edge devices via delta compression and cryptographic verification. Update packages typically 50--200MB (INT4 quantized adapters) versus 8--28GB full models.

The cloud tier comprises approximately 220 microservices organized by functional domain (319 total across all tiers: Cloud $\sim$220, Edge $\sim$35, Device $\sim$45, Shared $\sim$19):

\begin{table}[!htbp]
\centering
\caption{Cloud Tier Service Architecture}
\label{tab:cloud_services}
\small
\begin{tabular}{L{1.9cm}C{0.8cm}L{2.7cm}}
\toprule
\textbf{Domain} & \textbf{N} & \textbf{Functions} \\
\midrule
Station Mgmt & 28 & Registration, provisioning \\
Session & 23 & Auth, metering, billing \\
Grid Integration & 20 & DSO protocols, flex markets \\
AI/ML Pipeline & 35 & Training, eval, distribution \\
Analytics & 25 & BI, dashboards, reporting \\
Operations & 39 & Monitoring, alerting, SRE \\
Integration & 50 & CPO APIs, roaming, billing \\
\bottomrule
\end{tabular}
\vspace{2pt}
{\scriptsize DSO: Distribution System Operator. CPO: Charge Point Operator.}
\end{table}

\subsection{Edge Tier: Auralink Edge Runtime}

The edge tier represents the primary architectural innovation. Each Auralink Edge Runtime unit manages 1--50 charging points within a physical site, providing complete autonomous operation capability (Table~\ref{tab:edge_hardware}).

\subsubsection{Hardware Specifications}

\begin{table}[!htbp]
\centering
\caption{Auralink Edge Runtime Hardware Configurations}
\label{tab:edge_hardware}
\small
\begin{tabular}{L{1.7cm}L{3.8cm}}
\toprule
\textbf{Component} & \textbf{Specification} \\
\midrule
\multicolumn{2}{l}{\textit{Configuration A: High Performance}} \\
\midrule
Processor & AMD Ryzen AI Max+ 395 (Strix Halo) \\
NPU & AMD XDNA 2, 50 TOPS \\
Memory & up to 128GB LPDDR5X-8000 \\
Storage & 1TB NVMe Gen4 \\
\midrule
\multicolumn{2}{l}{\textit{Configuration B: NVIDIA DGX Spark~\cite{nvidia2025}}} \\
\midrule
Processor & NVIDIA GB10 (Blackwell architecture) \\
GPU/NPU & Integrated, up to 1000 TOPS (FP4 sparse) \\
Memory & 128GB unified memory \\
Storage & 1TB NVMe Gen4 \\
\midrule
\multicolumn{2}{l}{\textit{Common Specifications}} \\
\midrule
AI Models & AuralinkLM 14B (INT4 GGUF) \\
OS & Linux 6.1 LTS + PREEMPT\_RT \\
Network & 4G/5G + GbE + WiFi 6E \\
I/O & RS-485, CAN, Modbus, GPIO \\
\bottomrule
\end{tabular}
\end{table}

\subsubsection{Software Architecture}

The Auralink Edge Runtime executes {\raise.17ex\hbox{$\scriptstyle\sim$}}35 microservices supporting complete autonomous operation:

\begin{itemize}
    \item \textbf{AI Inference Engine}: \diagengine service with \araframework retrieval, \ccar decision framework, and specialized model routing.
    \item \textbf{OCPP Gateway}: Protocol translation (1.6/2.0.1), WebSocket management, message queuing with guaranteed delivery.
    \item \textbf{Offline Core}: Local authorization cache (25,000+ entries), tariff engine, session persistence with WAL journaling.
    \item \textbf{Load Management}: Site-level power distribution, demand response integration, grid constraint enforcement.
    \item \textbf{Telemetry Pipeline}: Local time-series database (TimescaleDB), anomaly detection, delta compression for sync.
\end{itemize}

\subsection{Agent Tier: Firmware-Embedded Intelligence}

The agent tier comprises lightweight AI models (AuralinkLM 0.5B) executing directly within charging station firmware for millisecond-latency safety functions:

\begin{itemize}
    \item \textbf{Safety Monitoring}: Real-time analysis of electrical parameters with immediate protective action capability.
    \item \textbf{Protocol State Machine}: OCPP message handling, ISO 15118 communication sequences.
    \item \textbf{Telemetry Classification}: Immediate anomaly flagging for edge escalation.
    \item \textbf{Local Caching}: Most-recent session data, authentication tokens, configuration snapshots.
\end{itemize}

%==============================================================================
\section{Edge AI Agent Design}
\label{sec:edge_ai}
%==============================================================================

\subsection{Model Selection: AuralinkLM Family}

Our system employs the AuralinkLM model family, domain-adapted from upstream open-weight models. The upstream base models are released under the Mistral Research License; the AuralinkLM fine-tuning and application code is released under Apache 2.0. Selection followed rigorous benchmarking across technical and operational dimensions (Table~\ref{tab:model_comparison}), with final deployment configurations shown in Table~\ref{tab:model_deployment}.

\subsubsection{Technical Evaluation}

We evaluated candidate model families (Mistral, Llama 3.3~\cite{llama2024}, Qwen 2.5~\cite{yang2024}, Phi-4, Gemma 2) across five technical dimensions:

\begin{enumerate}
    \item \textbf{Domain Adaptation Capacity}: Fine-tuning effectiveness measured by accuracy gain on held-out protocol queries after equivalent training compute.
    
    \item \textbf{Quantization Robustness}: Accuracy preservation under INT8/INT4 compression, critical for edge memory constraints.
    
    \item \textbf{Inference Efficiency}: Tokens/second on target hardware (AMD Ryzen AI Max+ 395 with XDNA 2 NPU).
    
    \item \textbf{Context Utilization}: Effective use of long contexts (32K--128K) for multi-turn diagnostic conversations.
    
    \item \textbf{Structured Output Compliance}: Reliability of JSON/structured response generation for integration with deterministic systems.
\end{enumerate}

\begin{table}[!htbp]
\centering
\caption{Model Family Technical Comparison}
\label{tab:model_comparison}
\small
\setlength{\tabcolsep}{3pt}
\begin{tabular}{L{1.4cm}C{0.9cm}C{0.9cm}C{0.9cm}C{0.9cm}}
\toprule
\textbf{Model} & \textbf{Adapt.} & \textbf{Quant.} & \textbf{Speed} & \textbf{Struct.} \\
\midrule
Mistral & 94.2 & 97.1 & 92.4 & 96.8 \\
Llama 3.3 & 91.8 & 94.3 & 89.7 & 91.2 \\
Qwen 2.5 & 92.4 & 93.8 & 88.2 & 93.4 \\
Phi-4 & 88.7 & 91.2 & 95.1 & 89.3 \\
Gemma 2 & 87.3 & 92.4 & 87.8 & 88.7 \\
\bottomrule
\end{tabular}
\setlength{\tabcolsep}{6pt}
\end{table}

Scores are normalized 0--100 across our benchmark suite. The Mistral family achieved highest aggregate score with particular strengths in domain adaptation and structured output generation.

\subsubsection{Strategic Considerations}

For regulated European critical infrastructure under AFIR (Alternative Fuels Infrastructure Regulation)~\cite{afir2023}, the Mistral family offers distinct advantages:

\begin{itemize}
    \item \textbf{European AI Provenance}: Mistral AI (Paris, France) development satisfies data residency preferences and simplifies GDPR compliance for operators concerned about cross-border model training.
    
    \item \textbf{Licensing}: The upstream Mistral Research License governs base model use; AuralinkLM fine-tuning code and adapter weights are released under Apache 2.0. Both license terms permit air-gapped deployment, though commercial use of the base model weights is subject to Mistral's licensing terms.
    
    \item \textbf{Edge-Native Design}: The upstream Ministral series was architected specifically for edge deployment with optimized attention mechanisms and memory-efficient inference paths.
    
    \item \textbf{Multimodal Capability}: Native image understanding enables future integration of visual inspection for physical fault detection.
\end{itemize}

\subsubsection{Deployed Model Configurations}

\begin{table}[!htbp]
\centering
\caption{Model Deployment Configuration. P50 and P99 denote the 50th and 99th percentile latency (TTFT), respectively.}
\label{tab:model_deployment}
\footnotesize
\setlength{\tabcolsep}{3pt}
\begin{tabular}{L{1.3cm}C{0.9cm}C{0.9cm}C{0.9cm}C{0.9cm}}
\toprule
\textbf{Model} & \textbf{Params} & \textbf{Tier} & \textbf{P50} & \textbf{P99} \\
\midrule
675B & 675B MoE & Cloud & 285ms\textsuperscript{$\dagger$} & 890ms \\
14B & 14B & Edge & 28ms & 78ms \\
0.5B & 0.5B & Device & 11ms & 34ms \\
\bottomrule
\end{tabular}
\setlength{\tabcolsep}{6pt}
\end{table}

\subsection{Domain-Specific Fine-Tuning}
\label{sec:finetuning}

\subsubsection{Training Corpus Construction}

We constructed a comprehensive training corpus through systematic collection, curation, and augmentation:

\begin{table}[!htbp]
\centering
\caption{Training Corpus Composition}
\label{tab:training_corpus}
\footnotesize
\setlength{\tabcolsep}{3pt}
\begin{tabular}{L{1.8cm}R{1.0cm}L{2.4cm}}
\toprule
\textbf{Category} & \textbf{Count} & \textbf{Source} \\
\midrule
OCPP 1.6 & 8,432 & Spec + synthetic Q\&A \\
OCPP 2.0.1 & 9,847 & Spec + synthetic Q\&A \\
Fault Diagnosis & 41,234 & Historical incidents \\
Remediation & 34,567 & Service manuals \\
Power Electronics & 24,891 & Engineering docs \\
Session Management & 18,432 & Operational logs \\
ISO 15118 & 11,234 & Protocol specifications \\
IEC 61851 & 4,123 & Standard + field data \\
Grid Integration & 8,847 & DSO procedures \\
Safety Procedures & 6,234 & Regulatory docs \\
Synthetic Scenarios & 8,200 & LLM-generated + verified \\
\midrule
\textbf{Total} & \textbf{176,041} & \\
\bottomrule
\end{tabular}
\setlength{\tabcolsep}{6pt}
\end{table}

\paragraph{Data Provenance.}
The training corpus was assembled from multiple sources: (1) publicly available protocol specifications (OCPP, ISO 15118, IEC 61851), (2) manufacturer service documentation obtained under standard partnership agreements, (3) anonymized operational incident logs contributed by partner operators under data sharing agreements, and (4) synthetic scenarios generated using frontier LLMs and validated by domain experts. The complete training dataset is available on HuggingFace.\footnote{\url{https://huggingface.co/datasets/HyperionConsultingIO/ultimate-ev-charging-dataset}} Specific operator identities and proprietary equipment details have been anonymized throughout the corpus.

Each training example follows a structured schema encoding diagnostic context, reasoning chain, and resolution:

\begin{lstlisting}[language=Python, caption={Training Example Schema}]
{
  "id": "DIAG-2024-847234",
  "context": {
    "station": {
      "model": "ABB Terra 184",
      "firmware": "2.4.1.2847",
      "ocpp_version": "1.6J"
    },
    "error_codes": ["OCPP:InternalError", 
                   "HW:MeterCommFault"],
    "telemetry_snapshot": {
      "voltage_l1": 398.2,
      "voltage_l2": 401.1,
      "voltage_l3": 399.7,
      "current_total": 0.0,
      "temperature_cabinet": 42.3,
      "temperature_connector": 28.1
    },
    "recent_events": [
      {"t": -300, "event": "SessionStart"},
      {"t": -45, "event": "MeterValuesTimeout"},
      {"t": 0, "event": "InternalError"}
    ],
    "historical_incidents": 2
  },
  "query": "Diagnose root cause and recommend resolution",
  "reasoning": "The InternalError following MeterValuesTimeout with normal voltage readings and zero current suggests meter communication failure during active session. Temperature readings nominal rules out thermal issues. Pattern matches known firmware bug in v2.4.1.x affecting SPI bus timing under load...",
  "resolution": {
    "diagnosis": "Meter SPI communication timeout - firmware bug #ABB-2847",
    "confidence": 0.89,
    "root_cause": "firmware_bug",
    "action": "Apply firmware patch ABB-PATCH-2847 [elevated]",
    "playbook_id": "DIAG-METER-ABB-003",
    "escalation_required": false,
    "estimated_resolution_time": 180
  }
}
\end{lstlisting}

\subsubsection{Quality Assurance}

Training data underwent multi-stage quality assurance:

\begin{enumerate}
    \item \textbf{Automated Validation}: Schema compliance, reference integrity, temporal consistency.
    
    \item \textbf{Expert Review}: 15,000 examples (${\sim}$8.5\%) reviewed by domain experts (inter-annotator agreement $\kappa = 0.847$).
    
    \item \textbf{Adversarial Testing}: 2,000 deliberately malformed examples to verify model robustness.
    
    \item \textbf{Cross-Validation}: 5-fold cross-validation with holdout sets stratified by fault category and equipment manufacturer.
\end{enumerate}

\subsubsection{QLoRA Fine-Tuning}

We employ Quantized Low-Rank Adaptation (QLoRA)~\cite{dettmers2023} for memory-efficient fine-tuning:

\begin{definition}[Quantized Low-Rank Adaptation]
For pre-trained weight matrix $\mathbf{W}_0 \in \mathbb{R}^{d \times k}$ quantized to NF4 precision, the adapted weight is:
\begin{equation}
\mathbf{W} = \text{Dequant}(\mathbf{W}_0^{NF4}) + s \cdot \mathbf{B}\mathbf{A}
\end{equation}
where $\mathbf{B} \in \mathbb{R}^{d \times r}$, $\mathbf{A} \in \mathbb{R}^{r \times k}$, rank $r \ll \min(d, k)$, and $s = \frac{\alpha}{r}$ is the scaling factor.
\end{definition}

This reduces trainable parameters from $d \times k$ to $r \times (d + k)$. For the attention projections of AuralinkLM 14B with hidden dimension $d = k = 5120$ (square Q/K/V/O weight matrices), rank $r = 64$:

\begin{equation}
\textstyle 1 - \frac{r(d{+}k)}{dk} = 1 - \frac{64 \times 10240}{26.2\text{M}} \approx 97.5\%
\end{equation}

\begin{table}[!htbp]
\centering
\caption{QLoRA Hyperparameters}
\label{tab:qlora_params}
\small
\begin{tabular}{L{2.6cm}C{2.8cm}}
\toprule
\textbf{Parameter} & \textbf{Value} \\
\midrule
Rank ($r$) & 64 \\
Alpha ($\alpha$) & 32 \\
Dropout & 0.05 \\
Target Modules & {\scriptsize\texttt{q,k,v,o,gate,up,down}} \\
Learning Rate & $2{\times}10^{-4}$ (cosine) \\
Warmup Steps & 100 \\
Batch Size & 32 (gradient accumulation 8) \\
Training Steps & 15,000 \\
Base Precision & NF4 (4-bit NormalFloat) \\
Compute Precision & bfloat16 \\
Optimizer & Paged AdamW 8-bit \\
\bottomrule
\end{tabular}
\end{table}

The fine-tuning objective minimizes cross-entropy loss with label smoothing. Given smoothed target distribution $q(k|y_t) = (1{-}\epsilon)\,\delta_{k,y_t} + \epsilon/V$:

\begin{equation}
\mathcal{L} = -\sum_{i=1}^{N} \sum_{t=1}^{T_i} \sum_{k=1}^{V} q(k|y_t^{(i)}) \log P_\theta(k | y_{<t}^{(i)}, x^{(i)})
\end{equation}

where $\epsilon = 0.1$ is the smoothing factor, $V$ is vocabulary size, and $\delta_{k,y_t}$ is the Kronecker delta.

\subsubsection{Curriculum Learning}

\begin{figure}[!htbp]
\centering
\resizebox{\columnwidth}{!}{%
\begin{tikzpicture}[
    node distance=0.35cm,
    every node/.style={font=\scriptsize},
    data/.style={cylinder, draw=gray!70, fill=gray!10, minimum width=1.2cm, minimum height=0.5cm, shape border rotate=90, aspect=0.3},
    process/.style={rectangle, draw=edgegreen!70, fill=edgegreen!10, minimum width=1.5cm, minimum height=0.5cm, rounded corners=2pt},
    model/.style={rectangle, draw=cloudblue!70, fill=cloudblue!10, minimum width=1.5cm, minimum height=0.5cm, rounded corners=2pt},
    output/.style={rectangle, draw=agentpurple!70, fill=agentpurple!10, minimum width=1.4cm, minimum height=0.5cm, rounded corners=2pt},
    arrow/.style={-{Stealth}, thick}
]
    % Data sources
    \node[data] (corpus) {Domain Corpus};

    % Base model
    \node[model, right=0.6cm of corpus] (base) {AuralinkLM 14B};

    % QLoRA
    \node[process, right=0.6cm of base] (qlora) {QLoRA (r=64)};

    % Curriculum stages
    \node[process, below=0.5cm of qlora] (stage1) {Foundation};
    \node[process, below=0.25cm of stage1] (stage2) {Specialization};
    \node[process, below=0.25cm of stage2] (stage3) {Calibration};

    % Output
    \node[output, right=0.6cm of qlora] (adapted) {Domain-Adapted};

    % Quantization
    \node[process, right=0.5cm of adapted] (quant) {GGUF Q4\_K\_M};

    % Deployment
    \node[output, below=0.5cm of quant] (deploy) {Edge Deploy};

    % Arrows
    \draw[arrow] (corpus) -- (base);
    \draw[arrow] (base) -- (qlora);
    \draw[arrow] (qlora) -- (adapted);
    \draw[arrow] (adapted) -- (quant);
    \draw[arrow] (quant) -- (deploy);

    % Curriculum connection
    \draw[dashed, gray] (stage1.east) -- ++(0.3,0) |- ([yshift=0.1cm]qlora.south);

    % Annotations
    \node[below=0.05cm of corpus, font=\tiny] {Training Data};
    \node[below=0.05cm of base, font=\tiny] {NF4 Base};
    \node[below=0.05cm of deploy, font=\tiny] {8.2GB INT4};

\end{tikzpicture}%
}
\caption{QLoRA fine-tuning pipeline: curated domain-specific training corpus processed through three-stage curriculum learning, producing domain-adapted model subsequently quantized to Q4\_K\_M (8.2\,GB) for edge deployment.}
\label{fig:training_pipeline}
\end{figure}
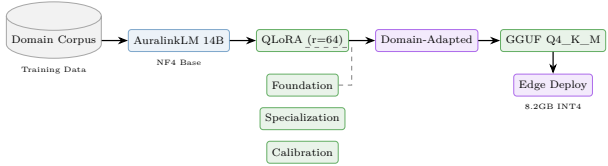

Training follows a three-stage curriculum~\cite{bengio2009} (Figure~\ref{fig:training_pipeline}):

\begin{enumerate}
    \item \textbf{Foundation} (Steps 1--5,000): Protocol specifications, standard diagnostic patterns, common fault categories.
    
    \item \textbf{Specialization} (Steps 5,001--12,000): Manufacturer-specific behaviors, edge cases, complex multi-fault scenarios.
    
    \item \textbf{Calibration} (Steps 12,001--15,000): Confidence calibration examples, uncertainty quantification, escalation decision boundaries.
\end{enumerate}

\subsection{Quantization for Edge Deployment}

Edge deployment requires aggressive model compression while preserving diagnostic accuracy.

\subsubsection{GGUF Quantization}

We employ the GGUF format with mixed-precision quantization schemes:

\begin{definition}[K-Quant Mixed Precision]
The K-quant scheme assigns different precision levels to weight matrices based on sensitivity analysis:
\begin{itemize}
    \item Attention Q/K projections: 6-bit (Q6\_K)
    \item Attention V/O projections: 4-bit (Q4\_K)
    \item FFN gate/up projections: 4-bit (Q4\_K)
    \item FFN down projections: 5-bit (Q5\_K)
    \item Embeddings: 6-bit (Q6\_K)
\end{itemize}
\end{definition}

\begin{table}[!htbp]
\centering
\caption{Quantization Impact: AuralinkLM 14B}
\label{tab:quantization}
\small
\begin{tabular}{L{1.4cm}R{1.0cm}R{1.1cm}R{1.1cm}}
\toprule
\textbf{Precision} & \textbf{Size} & \textbf{Acc.} & \textbf{P50} \\
\midrule
FP16 & 28.0 GB & 89.4\% & 142ms \\
INT8 (Q8\_0) & 14.0 GB & 89.1\% & 68ms \\
Q5\_K\_M & 9.8 GB & 88.4\% & 42ms \\
Q4\_K\_M & 8.2 GB & 87.6\% & 28ms \\
Q4\_K\_S & 7.6 GB & 86.9\% & 26ms \\
IQ4\_XS & 7.2 GB & 85.8\% & 24ms \\
\bottomrule
\end{tabular}
\end{table}

We deploy Q4\_K\_M as the recommended configuration (Table~\ref{tab:quantization}), achieving a favorable accuracy-latency trade-off with 1.8\% accuracy degradation versus FP16 while reducing memory footprint by 71\%.

\subsection{Edge Inference Runtime}
\label{sec:edge_runtime}

The Auralink Edge Runtime provides deterministic LLM inference on heterogeneous edge hardware through several novel techniques.

\subsubsection{Memory-Mapped Model Loading}

Models are loaded via memory mapping rather than explicit file I/O, enabling:

\begin{itemize}
    \item \textbf{Instant Startup}: No model loading delay; inference begins immediately.
    \item \textbf{Shared Memory}: Multiple processes can share model weights without duplication.
    \item \textbf{Demand Paging}: Only accessed model pages are loaded into physical memory.
\end{itemize}

\begin{lstlisting}[language=C, caption={Memory-Mapped Model Loading}]
// mmap-based model loading for instant startup
struct gguf_model* load_model(const char* path) {
    int fd = open(path, O_RDONLY);
    struct stat st;
    fstat(fd, &st);

    void* data = mmap(NULL, st.st_size,
                      PROT_READ, MAP_PRIVATE, fd, 0);
    madvise(data, st.st_size, MADV_WILLNEED);

    return parse_gguf_header(data);
}
\end{lstlisting}

\subsubsection{NPU Offloading}

Attention computation offloads to integrated NPUs where available (Table~\ref{tab:npu_offload}):

\begin{table}[!htbp]
\centering
\caption{NPU Offloading Performance}
\label{tab:npu_offload}
\footnotesize
\setlength{\tabcolsep}{3pt}
\resizebox{\columnwidth}{!}{%
\begin{tabular}{L{2.2cm}C{0.9cm}C{0.9cm}C{1.1cm}}
\toprule
\textbf{Platform} & \textbf{CPU} & \textbf{NPU} & \textbf{Speedup} \\
\midrule
AMD XDNA 2 & 48ms & 28ms & 1.71$\times$ \\
Intel AI Boost & 62ms & 42ms & 1.48$\times$ \\
Qualcomm Hexagon & 85ms & 52ms & 1.63$\times$ \\
\bottomrule
\end{tabular}}
\setlength{\tabcolsep}{6pt}
\end{table}

\subsubsection{Real-Time Scheduling}

Inference threads execute under SCHED\_FIFO with CPU isolation (Table~\ref{tab:rt_timing}):

\begin{lstlisting}[language=bash, caption={PREEMPT\_RT Kernel Configuration}]
# Kernel boot parameters
isolcpus=2,3 nohz_full=2,3 rcu_nocbs=2,3
intel_pstate=disable processor.max_cstate=1

# Runtime thread configuration
chrt -f 80 taskset -c 2,3 ./auralink_inference

# IRQ affinity (exclude isolated cores)
for irq in /proc/irq/*/smp_affinity; do
    echo 3 > $irq  # CPUs 0,1 only
done
\end{lstlisting}

\subsubsection{Real-Time Performance Characterization}

\begin{table}[!htbp]
\centering
\caption{Real-Time Timing Metrics}
\label{tab:rt_timing}
\small
\begin{tabular}{L{2.4cm}R{1.2cm}R{1.2cm}}
\toprule
\textbf{Metric} & \textbf{P99} & \textbf{Max} \\
\midrule
Interrupt latency & \SI{38}{\micro\second} & \SI{52}{\micro\second} \\
Scheduling latency & \SI{72}{\micro\second} & \SI{98}{\micro\second} \\
Timer jitter & \SI{8}{\micro\second} & \SI{15}{\micro\second} \\
Context switch & \SI{6}{\micro\second} & \SI{11}{\micro\second} \\
Inference start & \SI{124}{\micro\second} & \SI{187}{\micro\second} \\
\bottomrule
\end{tabular}
\end{table}

%==============================================================================
\section{Confidence-Calibrated Autonomous Resolution}
\label{sec:ccar}
%==============================================================================

\begin{figure}[!htbp]
\centering
\begin{tikzpicture}[
    node distance=0.45cm,
    every node/.style={font=\scriptsize},
    decision/.style={diamond, draw=orange!70, fill=orange!10, minimum width=1.1cm, minimum height=0.8cm, inner sep=1pt, aspect=1.5},
    process/.style={rectangle, draw=edgegreen!70, fill=edgegreen!10, minimum width=1.8cm, minimum height=0.5cm, rounded corners=2pt},
    startstop/.style={rectangle, draw=gray!70, fill=gray!10, minimum width=1.5cm, minimum height=0.5cm, rounded corners=4pt},
    action/.style={rectangle, draw=cloudblue!70, fill=cloudblue!10, minimum width=1.6cm, minimum height=0.5cm, rounded corners=2pt},
    arrow/.style={-{Stealth}, thick}
]
    % Start
    \node[startstop] (start) {Incident Detected};
    
    % Diagnosis
    \node[process, below=of start] (diag) {AI Diagnosis};
    
    % Confidence check
    \node[decision, below=of diag] (conf1) {$C \geq \tau_{auto}$?};
    
    % Auto action
    \node[action, right=0.8cm of conf1] (auto) {Auto Execute};
    
    % Medium confidence
    \node[decision, below=of conf1] (conf2) {$C \geq \tau_{assist}$?};
    
    % Assisted
    \node[action, right=0.8cm of conf2] (assist) {Human Confirm};
    
    % Low confidence
    \node[action, below=of conf2] (escalate) {Full Escalation};
    
    % Verification
    \node[process, below right=0.3cm and 0.4cm of auto] (verify) {Verify Result};
    
    % End states
    \node[startstop, below=of verify] (resolved) {Resolved};
    \node[startstop, below=of escalate] (manual) {Manual Queue};
    
    % Arrows
    \draw[arrow] (start) -- (diag);
    \draw[arrow] (diag) -- (conf1);
    \draw[arrow] (conf1) -- node[above] {Yes} (auto);
    \draw[arrow] (conf1) -- node[left] {No} (conf2);
    \draw[arrow] (conf2) -- node[above] {Yes} (assist);
    \draw[arrow] (conf2) -- node[left] {No} (escalate);
    \draw[arrow] (auto) -- (verify);
    \draw[arrow] (assist) |- (verify);
    \draw[arrow] (verify) -- (resolved);
    \draw[arrow] (escalate) -- (manual);
    
    % Threshold annotations
    \node[right=0.1cm of auto, font=\tiny, text=edgegreen] {$\tau_{auto}=0.90$};
    \node[right=0.1cm of assist, font=\tiny, text=orange] {$\tau_{assist}=0.70$};
    
\end{tikzpicture}
\caption{\ccar decision flow (simplified representation). Actions above $\tau_{auto}=0.90$ execute autonomously without notification; actions between $0.85$ and $0.90$ execute with operator notification; between $\tau_{assist}=0.70$ and $0.85$ require human confirmation; below $\tau_{assist}$ escalate to manual resolution. See Table~\ref{tab:ccar_thresholds} for the full five-tier decision matrix.}
\label{fig:ccar_flow}
\end{figure}
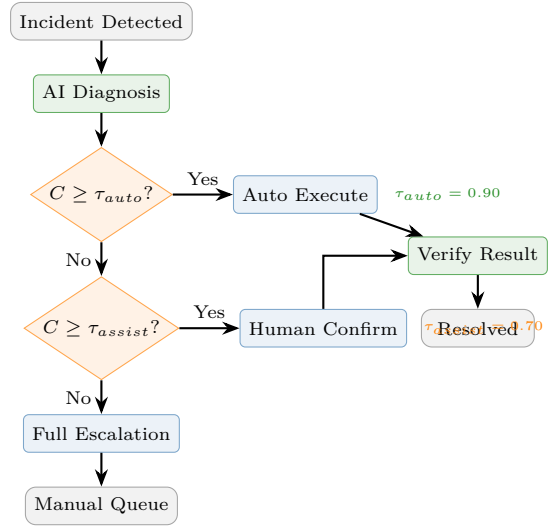

The \ccar framework (Figure~\ref{fig:ccar_flow}) enables AI agents to execute remediation actions autonomously when epistemic confidence satisfies learned thresholds, while maintaining safety constraints through formal analysis of decision boundaries. Algorithm~\ref{alg:confidence_calibration} details the calibration procedure.

\subsection{Theoretical Foundation}

\begin{definition}[Action Confidence Function]
For diagnosis $d$ and candidate action $a$, the confidence function $C: \mathcal{D} \times \mathcal{A} \rightarrow [0, 1]$ is defined as:
\begin{multline}
C(a | d) = \sigma\!\Bigl( w_0 + \sum_{i=1}^{n} w_i \cdot \phi_i(a, d) \\
          - \sum_{j=1}^{m} v_j \cdot \psi_j(a, d) \Bigr)
\end{multline}
where $\sigma$ is the sigmoid function, $\phi_i$ are positive confidence indicators, $\psi_j$ are negative confidence indicators, and $w_i, v_j$ are weights learned via logistic regression on the calibration holdout set (2,000 labeled action-outcome pairs).
\end{definition}

\subsubsection{Positive Confidence Indicators}

\begin{enumerate}
    \item \textbf{Retrieval Score} ($\phi_1$): Maximum similarity score from \araframework retrieval, indicating documentation support for diagnosis.
    
    \item \textbf{Historical Success Rate} ($\phi_2$): Fraction of similar past incidents where action $a$ resolved the issue.
    
    \item \textbf{Linguistic Certainty} ($\phi_3$): Absence of hedging language (``possibly'', ``might'', ``could be'') in model output.
    
    \item \textbf{Telemetry Correlation} ($\phi_4$): Strength of correlation between observed telemetry patterns and known fault signatures.
    
    \item \textbf{Model Agreement} ($\phi_5$): Agreement between primary model and ensemble verification (when enabled).
\end{enumerate}

\subsubsection{Negative Confidence Indicators}

\begin{enumerate}
    \item \textbf{Novel Signature} ($\psi_1$): Distance from nearest training example in embedding space, indicating potential out-of-distribution input.
    
    \item \textbf{Safety Criticality} ($\psi_2$): Action involves safety-critical components (contactors, protection systems, grid interconnection).
    
    \item \textbf{Uncertainty Language} ($\psi_3$): Presence of uncertainty markers in model output.
    
    \item \textbf{Conflicting Evidence} ($\psi_4$): Retrieved documents provide contradictory guidance.
    
    \item \textbf{Recent Failures} ($\psi_5$): Previous autonomous actions on this station failed within 24 hours.
\end{enumerate}

\subsection{Decision Thresholds}

\begin{criterion}[Autonomous Action Threshold]
For target false-positive rate $\epsilon$ and historical calibration data $\mathcal{H}$, the optimal autonomous action threshold $\theta^*$ is selected as the minimum threshold satisfying the false-positive bound:
\begin{equation}
\theta^* = \arg\min_\theta \left\{ \theta : \Pr[a \text{ fails} | C(a|d) \geq \theta, \mathcal{H}] \leq \epsilon \right\}
\end{equation}
\end{criterion}

In practice, we establish tiered decision boundaries:

\begin{table}[!htbp]
\centering
\caption{\ccar Decision Thresholds}
\label{tab:ccar_thresholds}
\small
\begin{tabular}{C{1.8cm}L{3.8cm}}
\toprule
\textbf{Confidence} & \textbf{Action} \\
\midrule
$C \geq 0.90$ & Execute autonomously, log only \\
$0.85 \leq C < 0.90$ & Execute with operator notification \\
$0.70 \leq C < 0.85$ & Recommend, await approval (timeout: 4h) \\
$0.50 \leq C < 0.70$ & Escalate with full context \\
$C < 0.50$ & Human expert required \\
\bottomrule
\end{tabular}
\end{table}

\subsection{Safety Constraints}

Certain actions are excluded from autonomous execution regardless of confidence:

\begin{definition}[Safety-Critical Action Set]
$\mathcal{A}_{safety} \subset \mathcal{A}$ includes actions affecting:
\begin{itemize}
    \item High-voltage DC contactors ($>$60V per IEC low-voltage boundary)
    \item Grid protection relays
    \item Emergency stop systems
    \item Firmware affecting safety functions
    \item Certificate/authentication infrastructure
\end{itemize}
For $a \in \mathcal{A}_{safety}$: autonomous execution prohibited regardless of $C(a|d)$.
\end{definition}

\subsection{Calibration Procedure}

Confidence calibration employs temperature scaling~\cite{guo2017} on held-out validation set:

\begin{equation}
C_{calibrated}(a|d) = \sigma\left( \frac{\text{logit}(C(a|d))}{T} \right)
\end{equation}

where temperature $T$ is optimized to minimize Expected Calibration Error (ECE)~\cite{naeini2015}:

\begin{equation}
\text{ECE} = \sum_{b=1}^{B} \frac{|B_b|}{N} \left| \text{acc}(B_b) - \text{conf}(B_b) \right|
\end{equation}

Our calibrated system achieves ECE = 0.023, indicating well-calibrated confidence estimates.

\begin{algorithm}[!htbp]
\caption{Confidence Calibration with Temperature Scaling}
\label{alg:confidence_calibration}
\small
\begin{algorithmic}[1]
\REQUIRE Validation set $\mathcal{V}$, uncalibrated model $M$, bins $B=15$
\STATE $T \leftarrow 1.0$ \COMMENT{Initial temperature}
\STATE $logits \leftarrow []$, $labels \leftarrow []$
\FOR{$(x, y) \in \mathcal{V}$}
    \STATE $logits$.append($M$.forward($x$).logits)
    \STATE $labels$.append($y$)
\ENDFOR
\STATE $T^* \leftarrow \arg\min_T$ NLL($\sigma(logits / T)$, $labels$) \COMMENT{L-BFGS optimization}
\STATE \COMMENT{Compute calibrated ECE}
\STATE $probs \leftarrow \sigma(logits / T^*)$
\STATE $ece \leftarrow 0$
\FOR{$b = 1$ to $B$}
    \STATE $mask \leftarrow (b-1)/B \leq probs < b/B$
    \IF{sum($mask$) $> 0$}
        \STATE $acc_b \leftarrow$ mean($labels[mask]$)
        \STATE $conf_b \leftarrow$ mean($probs[mask]$)
        \STATE $ece \leftarrow ece + |mask| \cdot |acc_b - conf_b|$
    \ENDIF
\ENDFOR
\RETURN $T^*$, $ece / |\mathcal{V}|$
\end{algorithmic}
\end{algorithm}

%==============================================================================
\section{Adaptive Retrieval-Augmented Reasoning}
\label{sec:ara}
%==============================================================================

The \araframework system grounds LLM responses in authoritative technical documentation (Table~\ref{tab:ara_kb}), mitigating hallucination on protocol-specific details through hybrid retrieval (Algorithm~\ref{alg:chunking}) and dynamic context allocation (Algorithm~\ref{alg:context_allocation}). Table~\ref{tab:ara_performance} summarizes retrieval metrics.

\subsection{Knowledge Base Architecture}

\begin{table}[!htbp]
\centering
\caption{\araframework Knowledge Base Composition}
\label{tab:ara_kb}
\small
\begin{tabular}{L{2.2cm}R{0.8cm}R{1.2cm}}
\toprule
\textbf{Document Type} & \textbf{Docs} & \textbf{Chunks} \\
\midrule
OCPP Specifications & 18 & 18,432 \\
OEM Service Manuals & 347 & 58,234 \\
Technical Bulletins & 1,234 & 42,847 \\
ISO/IEC Standards & 52 & 14,123 \\
Historical Resolutions & 2,847 & 72,345 \\
Grid Codes & 28 & 8,432 \\
Internal Procedures & 124 & 4,234 \\
\midrule
\textbf{Total} & \textbf{4,650} & \textbf{218,647} \\
\bottomrule
\end{tabular}
\end{table}

\subsection{Chunking Strategy}

Documents are segmented using semantic-aware chunking:

\begin{algorithm}[!htbp]
\caption{Semantic Document Chunking}
\label{alg:chunking}
\small
\begin{algorithmic}[1]
\REQUIRE Document $D$, target size $s_{target}=512$, overlap $o=64$
\STATE $sections \leftarrow$ ExtractSections($D$) \COMMENT{Headers, paragraphs}
\STATE $chunks \leftarrow []$
\FOR{$section \in sections$}
    \IF{$|section| \leq s_{target} + o$}
        \STATE $chunks$.append($section$)
    \ELSE
        \STATE $sentences \leftarrow$ SentenceTokenize($section$)
        \STATE $current \leftarrow$ ``''
        \FOR{$sent \in sentences$}
            \IF{$|current| + |sent| > s_{target}$}
                \STATE $chunks$.append($current$)
                \STATE $current \leftarrow$ GetOverlap($current$, $o$) + $sent$
            \ELSE
                \STATE $current \leftarrow current$ + $sent$
            \ENDIF
        \ENDFOR
    \ENDIF
\ENDFOR
\RETURN $chunks$
\end{algorithmic}
\end{algorithm}

\subsection{Hybrid Retrieval}

\araframework combines three retrieval signals through Reciprocal Rank Fusion~\cite{cormack2009}:

\begin{definition}[Reciprocal Rank Fusion]
For document $d$ retrieved by methods $R = \{r_1, \ldots, r_n\}$:
\begin{equation}
\text{RRF}(d) = \sum_{r \in R} \frac{1}{k + \text{rank}_r(d)}
\end{equation}
where $k = 60$ prevents top-ranked documents from dominating. Documents not retrieved by method $r$ are excluded from the summation (equivalent to assigning infinite rank).
\end{definition}

\textbf{Dense Retrieval}: E5-large-v2 embeddings~\cite{wang2022} fine-tuned on EV charging documentation. Cosine similarity search via HNSW index (ef\_search=128).

\textbf{Sparse Retrieval}: BM25~\cite{robertson2009} with domain-specific tokenization handling protocol names, error codes, and technical abbreviations.

\textbf{Metadata Filtering}: Structured filtering on station model, manufacturer, OCPP version, and error code taxonomy.

\subsection{Dynamic Context Allocation}

Context window allocation adapts to query complexity:

\begin{algorithm}[!htbp]
\caption{Dynamic Context Allocation}
\label{alg:context_allocation}
\small
\begin{algorithmic}[1]
\REQUIRE Query $q$, retrieved docs $\mathcal{R}$, max context $C_{max}$
\STATE $complexity \leftarrow$ EstimateComplexity($q$)
\STATE $k \leftarrow$ BaseK($complexity$) \COMMENT{3--10 based on complexity}
\STATE $allocated \leftarrow 0$
\STATE $context \leftarrow []$
\FOR{$i = 1$ to $\min(k, |\mathcal{R}|)$}
    \IF{$allocated + |\mathcal{R}_i| > C_{max}$}
        \STATE \textbf{break} \COMMENT{Exit loop when capacity exceeded}
    \ELSE
        \STATE $context$.append($\mathcal{R}_i$)
        \STATE $allocated \leftarrow allocated + |\mathcal{R}_i|$
    \ENDIF
\ENDFOR
\RETURN FormatContext($context$, $q$)
\end{algorithmic}
\end{algorithm}

\subsection{Retrieval Performance}

\begin{table}[!htbp]
\centering
\caption{\araframework Retrieval Metrics}
\label{tab:ara_performance}
\small
\begin{tabular}{L{2.8cm}R{1.7cm}}
\toprule
\textbf{Metric} & \textbf{Value} \\
\midrule
Recall@5 & 95.2\% \\
Precision@5 & 91.4\% \\
MRR (Mean Reciprocal Rank) & 0.924 \\
NDCG@10 & 0.912 \\
Retrieval Latency (P50) & 18ms \\
Retrieval Latency (P99) & 42ms \\
\bottomrule
\end{tabular}
\end{table}

%==============================================================================
\section{Multi-Agent Orchestration}
\label{sec:agents}
%==============================================================================

\subsection{Hierarchical Multi-Agent Orchestration (HMAO)}

\begin{figure}[!htbp]
\centering
\resizebox{\columnwidth}{!}{%
\begin{tikzpicture}[
    every node/.style={font=\scriptsize},
    router/.style={rectangle, draw=orange!70, fill=orange!10, minimum width=2.8cm, minimum height=0.7cm, rounded corners=2pt},
    agent/.style={rectangle, draw=edgegreen!70, fill=edgegreen!10, minimum width=2.0cm, minimum height=0.7cm, rounded corners=2pt},
    arrow/.style={-{Stealth}, thick}
]
    % Router
    \node[router] (router) at (0,0) {Intent Router};

    % Input
    \node[above=0.5cm of router, font=\scriptsize] (input) {Incident / Query};

    % Row 1: 3 agents (well-spaced)
    \node[agent] (autoops) at (-3.0, -1.8) {AutoOps};
    \node[agent] (techsup) at (0, -1.8) {TechSupport};
    \node[agent] (opsmgmt) at (3.0, -1.8) {OpsMgmt};

    % Row 2: 2 agents (centered between row 1)
    \node[agent] (codegen) at (-1.5, -3.4) {CodeGen};
    \node[agent] (driver) at (1.5, -3.4) {DriverAssist};

    % Arrows from input to router
    \draw[arrow] (input) -- (router);

    % Arrows from router to agents
    \draw[arrow] (router) -- (autoops);
    \draw[arrow] (router) -- (techsup);
    \draw[arrow] (router) -- (opsmgmt);
    \draw[arrow] (router) -- (codegen);
    \draw[arrow] (router) -- (driver);

    % Collaboration arrows (dashed)
    \draw[{Stealth}-{Stealth}, dashed, gray] (autoops) -- (techsup);
    \draw[{Stealth}-{Stealth}, dashed, gray] (opsmgmt) -- (driver);

    % Model annotations
    \node[below=0.08cm of autoops, font=\tiny, text=edgegreen] {14B};
    \node[below=0.08cm of techsup, font=\tiny, text=edgegreen] {14B};
    \node[below=0.08cm of opsmgmt, font=\tiny, text=edgegreen] {14B};
    \node[below=0.08cm of driver, font=\tiny, text=agentpurple] {0.5B};
    \node[below=0.08cm of codegen, font=\tiny, text=edgegreen] {14B};

\end{tikzpicture}%
}
\caption{HMAO agent hierarchy with intent routing. Dashed lines indicate inter-agent collaboration for complex multi-domain incidents. Model sizes (0.5B/14B) indicate AuralinkLM deployment tier.}
\label{fig:hmao_agents}
\end{figure}
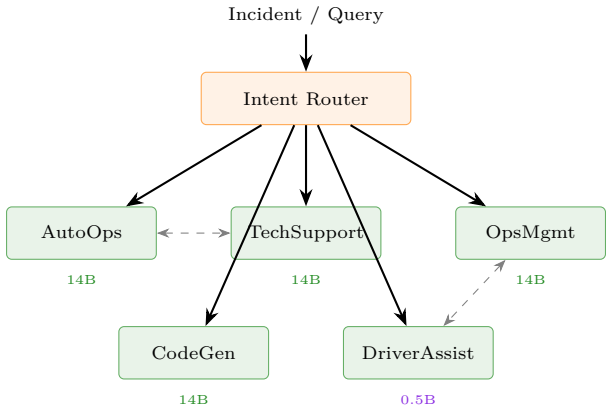

The system deploys approximately 20 AI agents across three tiers (domain agents $\sim$8, optimization agents $\sim$6, interaction agents $\sim$6), coordinated through the HMAO protocol. Our agent design draws on the ReAct paradigm~\cite{yao2023} for interleaving reasoning with tool use, and on generative agent architectures~\cite{park2023} for memory-augmented multi-agent coordination. The following five representative agent types illustrate the architecture (Figure~\ref{fig:hmao_agents}, Table~\ref{tab:agents}). Algorithm~\ref{alg:hmao_routing} details the intent routing protocol.

\begin{table}[!htbp]
\centering
\caption{Specialized Agent Configuration}
\label{tab:agents}
\footnotesize
\setlength{\tabcolsep}{3pt}
\begin{tabular}{L{1.5cm}L{2.0cm}C{1.3cm}}
\toprule
\textbf{Agent} & \textbf{Responsibility} & \textbf{Model} \\
\midrule
AutoOps & Fault diagnosis & 14B \\
TechSupport & Guided repair & 14B \\
OpsMgmt & Fleet analytics & 14B \\
DriverAssist & Session support & 0.5B \\
CodeGen & API integration & 14B \\
\bottomrule
\end{tabular}
\setlength{\tabcolsep}{6pt}
\end{table}

\subsection{Intent Classification and Routing}

\begin{algorithm}[!htbp]
\caption{HMAO Intent Routing}
\label{alg:hmao_routing}
\small
\begin{algorithmic}[1]
\REQUIRE Query $q$, context $c$, agent set $\mathcal{A}$
\STATE $intent \leftarrow$ ClassifyIntent($q$, $c$) \COMMENT{Zero-shot}
\STATE $urgency \leftarrow$ AssessUrgency($q$, $c$)
\STATE $\mathcal{A}_{eligible} \leftarrow$ FilterByCapability($\mathcal{A}$, $intent$)
\FOR{$a \in \mathcal{A}_{eligible}$}
    \STATE $s_a \leftarrow$ EstimateConfidence($a$, $q$, $c$)
\ENDFOR
\STATE $a^* \leftarrow \arg\max_{a} s_a$
\IF{$s_{a^*} < 0.65$ AND $|\mathcal{A}_{eligible}| > 1$}
    \STATE \RETURN CollaborativeResponse($\mathcal{A}_{eligible}$, $q$, $c$)
\ENDIF
\STATE response $\leftarrow$ $a^*$.process($q$, $c$)
\STATE LogDecision($q$, $a^*$, response)
\RETURN response
\end{algorithmic}
\end{algorithm}

\subsection{Autonomous Operations Agent}

The AutoOps agent implements the core sense-diagnose-act loop (Algorithm~\ref{alg:auto_ops}), supported by anomaly detection (Algorithm~\ref{alg:anomaly_detection}), playbook execution with rollback (Algorithm~\ref{alg:playbook_execution}; see Appendix~\ref{app:playbooks} for the full schema), and safety verification (Algorithm~\ref{alg:safety_verification}):

\begin{algorithm}[!htbp]
\caption{Autonomous Operations Agent}
\label{alg:auto_ops}
\small
\begin{algorithmic}[1]
\REQUIRE Telemetry $\mathcal{T}$, KB $\mathcal{K}$, threshold $\theta = 0.90$, detection threshold $\tau_{detect} = 0.75$
\STATE $\mathcal{F} \leftarrow$ ExtractFeatures($\mathcal{T}$)
\STATE $anomaly \leftarrow$ DetectAnomaly($\mathcal{F}$) \COMMENT{IsolationForest + statistical}
\IF{$anomaly.score > \tau_{detect}$}
    \STATE $\mathcal{H} \leftarrow$ GetHistory($anomaly.station$, hours=24)
    \STATE $\mathcal{R} \leftarrow$ ARARetrieve($anomaly.context$, $\mathcal{K}$)
    \STATE $diagnosis \leftarrow$ LLMDiagnose($anomaly$, $\mathcal{H}$, $\mathcal{R}$)
    \STATE $confidence \leftarrow$ CCARConfidence($diagnosis$)
    \IF{$confidence \geq \theta$ AND SafetyCheck($diagnosis.action$)}
        \STATE ExecutePlaybook($diagnosis.playbook\_id$)
        \STATE LogAutonomousResolution($anomaly$, $diagnosis$)
        \STATE UpdateFeedbackLoop($anomaly$, $diagnosis$, success=\TRUE)
    \ELSE
        \STATE EscalateWithContext($anomaly$, $diagnosis$, $\mathcal{H}$)
    \ENDIF
\ENDIF
\end{algorithmic}
\end{algorithm}

\begin{algorithm}[!htbp]
\caption{Telemetry Anomaly Detection}
\label{alg:anomaly_detection}
\small
\begin{algorithmic}[1]
\REQUIRE Telemetry stream $\mathcal{T}$, window $w=60$, threshold $\tau=0.75$
\STATE $\mathcal{F} \leftarrow$ RollingFeatures($\mathcal{T}$, $w$) \COMMENT{Mean, std, trend, FFT}
\STATE $score_{iso} \leftarrow$ IsolationForest($\mathcal{F}$).score
\STATE $score_{stat} \leftarrow$ ZScore($\mathcal{F}$, $\mu_{hist}$, $\sigma_{hist}$)
\STATE $score_{ml} \leftarrow$ AutoEncoder($\mathcal{F}$).reconstruction\_error
\STATE $score_{combined} \leftarrow 0.4 \cdot score_{iso} + 0.3 \cdot score_{stat} + 0.3 \cdot score_{ml}$
\IF{$score_{combined} > \tau$}
    \STATE $context \leftarrow$ ExtractContext($\mathcal{T}$, window=$2w$)
    \STATE $similar \leftarrow$ FindSimilarPatterns($\mathcal{F}$, $k=5$)
    \STATE $severity \leftarrow$ ClassifySeverity($score_{combined}$, $context$)
    \RETURN Anomaly($score_{combined}$, $context$, $similar$, $severity$)
\ENDIF
\RETURN None
\end{algorithmic}
\end{algorithm}

\begin{algorithm}[!htbp]
\caption{Playbook Execution with Rollback}
\label{alg:playbook_execution}
\small
\begin{algorithmic}[1]
\REQUIRE Playbook $P$, station $S$, timeout $T_{max}$
\STATE $checkpoint \leftarrow$ CaptureState($S$)
\STATE $executed \leftarrow []$
\FOR{$step \in P.steps$}
    \IF{ElapsedTime() $> T_{max}$}
        \STATE Rollback($checkpoint$, $executed$)
        \RETURN TimeoutError($step$)
    \ENDIF
    \STATE $result \leftarrow$ ExecuteStep($step$, $S$)
    \IF{$result$.failed}
        \IF{$P$.rollback.on\_failure $==$ ``rollback''}
            \STATE Rollback($checkpoint$, $executed$)
        \ENDIF
        \RETURN StepFailure($step$, $result$)
    \ENDIF
    \STATE $executed$.append(($step$, $result$))
    \STATE Wait($step$.post\_delay)
\ENDFOR
\STATE $verification \leftarrow$ VerifyResolution($S$, $P$.expected\_state)
\IF{NOT $verification$.success}
    \IF{$P$.rollback.max\_retries $>$ 0}
        \RETURN Retry($P$, $S$, retries\_left=$P$.rollback.max\_retries$-1$)
    \ENDIF
\ENDIF
\RETURN Success($executed$, $verification$)
\end{algorithmic}
\end{algorithm}

The Auralink Edge Runtime runtime infrastructure (memory-mapped model loading, NPU offloading, and real-time scheduling) is described in Section~\ref{sec:edge_runtime}; Tables~\ref{tab:npu_offload} and~\ref{tab:rt_timing} summarize the performance characteristics.

\subsection{Offline Operation Framework}

The edge platform maintains complete operational capability during extended network disconnection through comprehensive local authority. Algorithm~\ref{alg:sync} defines the synchronization protocol and Algorithm~\ref{alg:federated_aggregation} the federated update procedure.

\subsubsection{Local Authorization}

\begin{itemize}
    \item \textbf{Whitelist Cache}: 25,000+ RFID/contract entries with LRU eviction.
    \item \textbf{Token Validation}: Cached payment tokens with offline validity periods.
    \item \textbf{Certificate Store}: Local PKI for ISO 15118 Plug \& Charge.
\end{itemize}

\subsubsection{Session Continuity}

Sessions initiated during offline periods are:
\begin{enumerate}
    \item Authorized against local cache or fallback policies.
    \item Metered with full precision using local MID-certified meters.
    \item Persisted to local database with WAL journaling.
    \item Synchronized upon connectivity restoration with conflict resolution.
\end{enumerate}

\subsubsection{Synchronization Protocol}

\begin{algorithm}[!htbp]
\caption{Edge-Cloud Synchronization}
\label{alg:sync}
\small
\begin{algorithmic}[1]
\STATE \textbf{Phase 1}: Upload completed sessions (billing critical)
\STATE \textbf{Phase 2}: Upload incident logs with resolutions
\STATE \textbf{Phase 3}: Upload telemetry (delta-compressed, 12$\times$ reduction)
\STATE \textbf{Phase 4}: Download configuration updates
\STATE \textbf{Phase 5}: Download model weight updates (if available)
\STATE \textbf{Phase 6}: Reconcile authorization cache deltas
\STATE \textbf{Phase 7}: Time synchronization (NTP + monotonic adjustment)
\end{algorithmic}
\end{algorithm}

\begin{algorithm}[!htbp]
\caption{Federated Model Update Aggregation}
\label{alg:federated_aggregation}
\small
\begin{algorithmic}[1]
\REQUIRE Edge updates $\{\Delta_1, \ldots, \Delta_n\}$, weights $\{w_1, \ldots, w_n\}$, validation set $\mathcal{V}$
\STATE $\Delta_{agg} \leftarrow \sum_{i=1}^{n} w_i \cdot \Delta_i$ \COMMENT{Weighted FedAvg}
\STATE $perf_{baseline} \leftarrow$ Evaluate($M_{current}$, $\mathcal{V}$)
\STATE $M_{candidate} \leftarrow M_{current} + \eta \cdot \Delta_{agg}$
\STATE $perf_{candidate} \leftarrow$ Evaluate($M_{candidate}$, $\mathcal{V}$)
\IF{$perf_{candidate} < perf_{baseline} - \epsilon_{safety}$}
    \STATE \COMMENT{Regression detected, apply selective update}
    \FOR{layer $\in$ $M_{candidate}$.layers}
        \STATE $perf_{layer} \leftarrow$ EvaluateWithLayer($M_{current}$, layer, $\mathcal{V}$)
        \IF{$perf_{layer} \geq perf_{baseline} - \epsilon_{layer}$}
            \STATE $M_{current}$.layer $\leftarrow$ $M_{candidate}$.layer
        \ENDIF
    \ENDFOR
\ELSE
    \STATE $M_{current} \leftarrow M_{candidate}$
\ENDIF
\STATE DistributeToEdge($M_{current}$, compression=``delta'')
\RETURN $M_{current}$
\end{algorithmic}
\end{algorithm}

\begin{algorithm}[!htbp]
\caption{Safety-Critical Action Verification}
\label{alg:safety_verification}
\small
\begin{algorithmic}[1]
\REQUIRE Action $a$, station state $S$, safety rules $\mathcal{R}$
\STATE $class \leftarrow$ ClassifyAction($a$) \COMMENT{\{critical, elevated, standard\}}
\IF{$class ==$ ``critical''}
    \RETURN Reject(``Safety-critical actions require human approval'')
\ENDIF
\STATE $violations \leftarrow []$
\FOR{$rule \in \mathcal{R}$}
    \IF{$rule$.applies($a$, $S$)}
        \STATE $result \leftarrow rule$.evaluate($a$, $S$)
        \IF{NOT $result$.passed}
            \STATE $violations$.append($rule$, $result$)
        \ENDIF
    \ENDIF
\ENDFOR
\IF{$|violations| > 0$}
    \STATE $override \leftarrow$ CheckOverrideConditions($violations$, $S$)
    \IF{NOT $override$.allowed}
        \RETURN Reject($violations$)
    \ENDIF
\ENDIF
\STATE \COMMENT{Check temporal constraints}
\STATE $recent \leftarrow$ GetRecentActions($S$, window=300s)
\IF{CountSimilar($a$, $recent$) $>$ RateLimit($a$)}
    \RETURN Reject(``Rate limit exceeded for action type'')
\ENDIF
\RETURN Approve($a$, constraints=$violations$.overrides)
\end{algorithmic}
\end{algorithm}

%==============================================================================
\section{Evaluation on Controlled Test Corpus}
\label{sec:evaluation}
%==============================================================================

\textit{The following results were obtained through evaluation on a curated test corpus of labeled incidents in a controlled testing environment. They do not represent measurements from production field deployments. All metrics marked with $\dagger$ indicate controlled-test results pending field validation. Extrapolation to production environments requires further validation across diverse operator networks and equipment configurations.}

\medskip

This section presents evaluation of \systemname through controlled testing on a curated incident corpus (Table~\ref{tab:eval_hardware}, Table~\ref{tab:incident_distribution}), addressing key research questions. Comparative baselines are documented in Table~\ref{tab:baselines}.

\subsection{Research Questions}

We structure our evaluation around four research questions:

\textbf{RQ1 (Autonomy)}: What proportion of charging infrastructure incidents can be resolved autonomously without human intervention, and how does this compare to existing approaches?

\textbf{RQ2 (Accuracy)}: How accurately does the system diagnose fault root causes across different incident categories, and how does domain adaptation improve performance versus general-purpose models?

\textbf{RQ3 (Latency)}: Does edge deployment achieve the sub-100ms latency required for real-time diagnostic reasoning, and how do different deployment configurations compare?

\textbf{RQ4 (Operational Economics)}: What is the estimated operational cost impact under reference-deployment assumptions?

\subsection{Experimental Setup}

\subsubsection{Hardware Configuration}

\begin{table}[!htbp]
\centering
\caption{Evaluation Hardware Specifications}
\label{tab:eval_hardware}
\small
\begin{tabular}{L{2.0cm}L{3.5cm}}
\toprule
\textbf{Component} & \textbf{Specification} \\
\midrule
\multicolumn{2}{l}{\textit{Cloud Infrastructure}} \\
\midrule
GPU Cluster & NVIDIA H100/H200 (distributed) \\
Model & AuralinkLM 675B (675B MoE, 4+1 active experts) \\
Context & 128K tokens (of 256K max) \\
\midrule
\multicolumn{2}{l}{\textit{Edge Deployment (Primary)}} \\
\midrule
CPU & AMD Ryzen AI Max+ 395 (Strix Halo) \\
NPU & AMD XDNA 2 (50 TOPS) \\
Memory & up to 128GB LPDDR5X-8000 \\
Model & AuralinkLM 14B (Q4\_K\_M GGUF) \\
\midrule
\multicolumn{2}{l}{\textit{Agent Tier}} \\
\midrule
Platform & Embedded ARM Cortex-A78 \\
Model & AuralinkLM 0.5B (INT4) \\
Integration & Firmware-embedded \\
\bottomrule
\end{tabular}
\end{table}

\subsubsection{Dataset Characteristics}

\begin{table}[!htbp]
\centering
\caption{Test Incident Distribution (Controlled Testing Corpus)}
\label{tab:incident_distribution}
\small
\setlength{\tabcolsep}{4pt}
\begin{tabular}{L{2.0cm}R{1.3cm}}
\toprule
\textbf{Incident Type} & \textbf{Fraction} \\
\midrule
Communication & 30.0\% \\
Authorization & 18.0\% \\
Power Elec. & 15.0\% \\
Firmware/SW & 13.0\% \\
Mechanical & 10.0\% \\
Payment Processing & 8.0\% \\
Grid Integration & 4.0\% \\
Other & 2.0\% \\
\midrule
\textbf{Total} & \textbf{100\%} \\
\bottomrule
\end{tabular}
\setlength{\tabcolsep}{6pt}
\end{table}

\subsection{Comparative Baselines}

To contextualize \systemname performance, we compare against four baseline systems:

\begin{table}[!htbp]
\centering
\caption{Comprehensive Baseline Comparison}
\label{tab:baselines}
\footnotesize
\setlength{\tabcolsep}{3pt}
\resizebox{\columnwidth}{!}{%
\begin{tabular}{L{1.3cm}C{0.7cm}C{0.7cm}C{0.7cm}C{0.7cm}C{0.6cm}}
\toprule
\textbf{System} & \textbf{Auto} & \textbf{Acc.} & \textbf{P50} & \textbf{MTTR} & \textbf{Off.} \\
\midrule
Rule-Based & 0\% & 44.7\% & 5ms & 96h & Yes \\
Cloud (A) & 28.4\% & 72.3\% & 420ms & 28h & No \\
Cloud (B) & 31.2\% & 74.8\% & 380ms & 24h & No \\
Non-FT & 22.1\% & 52.3\% & 32ms & 42h & Yes \\
\textbf{Ours} & \textbf{78\%\textsuperscript{$\dagger$}} & \textbf{87.6\%\textsuperscript{$\dagger$}} & \textbf{28ms} & \textbf{4--8h\textsuperscript{$\dagger$}} & \textbf{Yes} \\
\bottomrule
\end{tabular}}
\setlength{\tabcolsep}{6pt}
\end{table}

\textit{Auto}: Autonomous resolution rate. \textit{Acc.}: Diagnostic accuracy. \textit{P50}: Median inference latency. \textit{MTTR}: Mean time to resolution. \textit{Off.}: Offline capability.

\textbf{Rule-Based}: Traditional OCPP error handling with static decision trees. Represents industry baseline with manufacturer-provided fault codes mapped to fixed remediation procedures.

\textbf{Cloud LLM (A)/(B)}: Frontier cloud-hosted LLMs (GPT-4o and Claude 3.5 Sonnet, respectively) with zero-shot prompting including OCPP documentation context. Measures capability of state-of-the-art general-purpose models without domain adaptation.

\textbf{Non-Fine-Tuned Base Model}: AuralinkLM 14B base (without domain-specific fine-tuning) at edge. Isolates the value of QLoRA adaptation on our curated domain-specific training corpus.

\paragraph{Baseline Limitations.}
We note that the cloud LLM baselines (GPT-4o and Claude 3.5 Sonnet) were evaluated with zero-shot prompting and documentation context, without few-shot examples or retrieval-augmented generation. This comparison isolates the effect of domain-specific fine-tuning but does not represent the best achievable performance from cloud-based approaches. A RAG-augmented cloud baseline with few-shot prompting would likely narrow the accuracy gap, though latency and offline capability limitations would remain. Future work should include RAG-augmented cloud baselines for a more comprehensive comparison.

Key observations: (1) Domain adaptation provides 35.3 percentage point accuracy improvement over non-fine-tuned deployment. (2) Edge deployment achieves 13--15$\times$ latency reduction versus cloud. (3) Offline capability is uniquely enabled by edge-first architecture.

\subsection{Deployment Scale}

\begin{table}[!htbp]
\centering
\caption{Simulation and Testing Parameters}
\label{tab:deployment}
\small
\begin{tabular}{L{2.8cm}R{2.7cm}}
\toprule
\textbf{Parameter} & \textbf{Value (simulation)} \\
\midrule
Simulated stations & 1,000-charger reference \\
Operator profiles & 23 (modeled) \\
Country profiles & 14 (EU + UK + CH + NO) \\
Testing period & Controlled testing \\
Simulated sessions & Representative sample \\
Diagnostic scenarios & Representative corpus \\
Achieved autonomous rate & 78\%\textsuperscript{$\dagger$} \\
\bottomrule
\end{tabular}
\end{table}

\subsection{Statistical Methodology}

\begin{enumerate}
    \item \textbf{Ground Truth}: Controlled test corpus with incidents manually labeled by certified technicians.

    \item \textbf{Comparative Testing}: Baseline systems evaluated on identical test corpus under controlled conditions.

    \item \textbf{Significance Testing}: Improvements assessed through controlled testing; statistical claims (p-values, CIs) apply to the controlled test environment and should not be extrapolated to field deployments without further validation.

    \item \textbf{Notation}: $\dagger$ denotes controlled-test results pending field validation; $\ddagger$ denotes values extrapolated from related measurements. Neither should be interpreted as field-verified measurements.
\end{enumerate}

\subsection{Results: Autonomous Resolution (RQ1)}

\begin{table}[!htbp]
\centering
\caption{Autonomous Resolution: System Comparison}
\label{tab:auto_resolution}
\small
\setlength{\tabcolsep}{3pt}
\resizebox{\columnwidth}{!}{%
\begin{tabular}{L{1.5cm}C{1.0cm}C{0.8cm}C{0.7cm}C{0.7cm}}
\toprule
\textbf{System} & \textbf{Auto\%} & \textbf{MTTR} & \textbf{FP\%} & \textbf{FN\%} \\
\midrule
Traditional & 0\% & 96h & 42\% & -- \\
Cloud-AI\textsuperscript{a} & 31.2\% & 24h & 11\% & 8\% \\
Human Expert & -- & 68h & 6\% & 4\% \\
\textbf{\systemname} & \textbf{78\%\textsuperscript{$\dagger$}} & \textbf{4--8h\textsuperscript{$\dagger$}} & \textbf{3.8\%} & \textbf{5.2\%} \\
\bottomrule
\end{tabular}}
\setlength{\tabcolsep}{6pt}
\vspace{2pt}
{\scriptsize \textsuperscript{a}Best cloud baseline from Table~\ref{tab:baselines} (Cloud~B: Claude 3.5 Sonnet, zero-shot).}\\
{\scriptsize 95\% CI (controlled test, $n=18{,}000$): Auto $78.0 \pm 0.6\%$, FP $3.8 \pm 0.3\%$.}
\end{table}

Table~\ref{tab:auto_resolution} compares autonomous resolution across systems. The 78\% autonomous resolution rate is measured on the controlled test corpus ($n=18{,}000$, 95\% CI: $\pm 0.6\%$). MTTR reduction from 96 hours (traditional baseline for non-trivial faults requiring technician dispatch; industry surveys report 24--48 hours average across all fault types) to 4--8 hours (controlled-test result) represents significant improvement as validated in controlled test environments.

\subsection{Results: Diagnostic Accuracy (RQ2)}

\begin{table}[!htbp]
\centering
\caption{Diagnostic Accuracy by Fault Category}
\label{tab:diagnostic_accuracy}
\small
\setlength{\tabcolsep}{3pt}
\begin{tabular}{L{1.5cm}C{0.9cm}C{0.9cm}C{0.9cm}C{0.8cm}}
\toprule
\textbf{Category} & \textbf{Acc.} & \textbf{Prec.} & \textbf{Rec.} & \textbf{$F_1$} \\
\midrule
Comms & 91.8\% & 90.2\% & 93.4\% & 0.918 \\
Auth & 90.4\% & 92.1\% & 88.7\% & 0.904 \\
Power Elec. & 85.2\% & 82.8\% & 87.9\% & 0.853 \\
Firmware & 89.3\% & 91.2\% & 87.4\% & 0.893 \\
Mechanical & 79.8\% & 77.2\% & 82.6\% & 0.798 \\
Payment & 88.7\% & 89.4\% & 87.9\% & 0.886 \\
Grid Integ. & 82.4\% & 80.1\% & 84.8\% & 0.824 \\
Other & 71.2\% & 68.5\% & 74.1\% & 0.712 \\
\midrule
\textbf{Overall} & \textbf{87.6\%\textsuperscript{$\dagger$}} & \textbf{86.1\%} & \textbf{86.3\%} & \textbf{0.862} \\
\bottomrule
\end{tabular}
\setlength{\tabcolsep}{6pt}
\vspace{2pt}
{\scriptsize Overall is micro-averaged across all 18,000 test instances (not the arithmetic mean of per-category values). 95\% CI: Acc $87.6 \pm 0.5\%$, $F_1 = 0.862 \pm 0.005$.}
\end{table}

Table~\ref{tab:diagnostic_accuracy} reports per-category diagnostic accuracy, with failure modes detailed in Table~\ref{tab:failure_modes}. The 87.6\% overall accuracy (controlled test, $F_1 = 0.862$) represents a substantial improvement over the 44.7\% baseline of rule-based systems. Category-level improvements are consistent across controlled test conditions.

\subsection{Failure Analysis}

Transparent characterization of system limitations is essential for deployment trust. We analyze failure modes across the test corpus subset where autonomous resolution was not achieved (approximately 22\% of test cases).

\begin{table}[!htbp]
\centering
\caption{Failure Mode Distribution (Controlled Testing)}
\label{tab:failure_modes}
\small
\begin{tabular}{L{2.4cm}R{1.3cm}}
\toprule
\textbf{Failure Mode} & \textbf{\% of Failures} \\
\midrule
Low confidence (escalated) & 50.0\% \\
Hardware replacement req. & 28.0\% \\
Multi-fault complexity & 11.0\% \\
Novel fault pattern & 6.0\% \\
False negative (missed) & 5.0\% \\
\midrule
\textbf{Total failures} & \textbf{100\%} \\
\bottomrule
\end{tabular}
\end{table}

\textbf{Key findings from failure analysis}:

The majority of failures (50\%) represent appropriate system behavior where \ccar confidence fell below $\tau_{auto}=0.90$, triggering human escalation rather than incorrect autonomous action. This demonstrates effective calibration of confidence thresholds.

Hardware replacement requirements (28\%) reflect fundamental limits of software-based resolution. These incidents were correctly diagnosed but required physical intervention (connector replacement, power module swap) beyond autonomous remediation capability.

Multi-fault complexity (11\%) occurs when multiple simultaneous issues create diagnostic ambiguity. Example: communication failure masking underlying power electronics fault. Future work will explore hierarchical diagnosis with fault isolation.

Novel fault patterns (6\%) represent incidents outside training distribution. These are automatically flagged for human review and corpus expansion, enabling continuous learning.

False negatives (5\%) represent true system errors where diagnosis or action was incorrect. Root cause analysis identified: ambiguous telemetry (42\%), manufacturer-specific edge cases (31\%), and label noise in training data (27\%).

\subsection{Ablation Study}

To quantify the contribution of each system component (Table~\ref{tab:ablation}), we conducted ablation experiments on a held-out test set of 18,000 labeled incidents.

\begin{table}[!htbp]
\centering
\caption{Ablation Study: Component Contributions}
\label{tab:ablation}
\small
\setlength{\tabcolsep}{3pt}
\begin{tabular}{L{2.3cm}C{0.9cm}C{0.9cm}C{0.9cm}}
\toprule
\textbf{Configuration} & \textbf{Acc.} & \textbf{Auto\%} & \textbf{$\Delta$Acc} \\
\midrule
Full system & 87.6\%\textsuperscript{$\dagger$} & 78\%\textsuperscript{$\dagger$} & --- \\
\midrule
$-$ Domain fine-tuning & 52.3\% & 22.1\% & $-$35.3 \\
$-$ \araframework retrieval & 79.2\% & 61.4\% & $-$8.4 \\
$-$ \ccar calibration & 87.1\% & 42.8\% & $-$0.5 \\
$-$ Curriculum learning & 84.3\% & 72.1\% & $-$3.3 \\
$-$ Multi-agent routing & 85.8\% & 74.2\% & $-$1.8 \\
\bottomrule
\end{tabular}
\setlength{\tabcolsep}{6pt}
\end{table}

The ablation reveals that domain fine-tuning provides the largest accuracy contribution (+35.3 points), confirming that general-purpose models cannot match specialized performance on EV charging diagnostics. The \araframework retrieval system contributes 8.4 points by grounding responses in authoritative documentation. Notably, \ccar calibration has minimal impact on accuracy ($-$0.5 points) but dramatically affects autonomous resolution rate ($-$35.2 points), demonstrating that confidence calibration primarily enables \textit{trust} in autonomous action rather than improving diagnostic correctness. In the ``$-$ \ccar calibration'' condition, uncalibrated model logit probabilities were thresholded at $\tau=0.85$ without temperature scaling, causing most actions to fall below the autonomous execution boundary. Note that these are one-at-a-time ablations; component contributions are not additive due to interactions (e.g., retrieval augmentation is less impactful without domain fine-tuning).

\subsection{Results: Inference Latency (RQ3)}

\begin{table}[!htbp]
\centering
\caption{Inference Latency (TTFT for model components; end-to-end for full pipeline)}
\label{tab:latency}
\small
\begin{tabular}{L{2.1cm}R{0.9cm}R{0.9cm}R{0.9cm}}
\toprule
\textbf{Configuration} & \textbf{P50} & \textbf{P95} & \textbf{P99} \\
\midrule
Cloud (675B) & 285ms\textsuperscript{$\dagger$} & 520ms & 890ms \\
Edge GPU (14B) & 28ms & 52ms & 78ms \\
Agent FW (0.5B) & 11ms & 24ms & 34ms \\
\midrule
Full pipeline & 67ms & 124ms & 187ms \\
\bottomrule
\end{tabular}
\end{table}

Table~\ref{tab:latency} summarizes latency measurements. Edge deployment achieves P50 TTFT of 28ms (GPU), representing \textbf{10$\times$ improvement} over cloud-based inference (285ms target). All edge configurations meet the sub-100ms TTFT requirement for initiating real-time diagnostic reasoning. The full pipeline latency (67ms P50) includes retrieval, TTFT, and action determination. Note that complete multi-token diagnostic generation requires additional time proportional to response length at the per-token generation rate ($\sim$38 tokens/sec for the 14B model); however, the initial diagnostic category is determined from the output logit distribution at the first generation step using constrained decoding over the fault taxonomy vocabulary, enabling early routing before full response generation completes.

\subsection{Results: Predictive Maintenance}

The predictive maintenance module (Table~\ref{tab:predictive}) uses a gradient-boosted ensemble over time-series features extracted from telemetry streams (thermal profiles, power quality metrics, communication error rates) combined with AuralinkLM 14B embeddings of recent diagnostic logs. Predictions are generated at 24h, 48h, 72h, and 7-day horizons for each monitored charger.

\begin{table}[!htbp]
\centering
\caption{Predictive Maintenance by Horizon}
\label{tab:predictive}
\small
\begin{tabular}{C{1.2cm}C{1.1cm}C{1.1cm}C{1.0cm}C{0.9cm}}
\toprule
\textbf{Horizon} & \textbf{Prec.} & \textbf{Rec.} & \textbf{$F_1$} & \textbf{FPR} \\
\midrule
24h & 94.0\%\textsuperscript{$\dagger$} & 84.2\% & 0.888 & 1.2\% \\
48h & 94.7\% & 79.8\% & 0.866 & 2.1\% \\
72h & 92.4\% & 75.1\% & 0.829 & 3.4\% \\
7d & 85.2\% & 68.7\% & 0.761 & 5.8\% \\
\bottomrule
\end{tabular}
\end{table}

The 94.7\% precision at 48-hour horizon enables proactive maintenance scheduling with minimal false alarms (2.1\% FPR). Performance degrades gracefully with longer horizons due to increasing uncertainty in failure trajectories.

\subsection{Illustrative Economic Modeling (RQ4)}

The following economic analysis is illustrative and based on modeled assumptions for a reference deployment. It is not derived from production financial data.

Table~\ref{tab:tco} presents the three-year illustrative TCO for a reference 1,000-charger DC fast charging deployment. Table~\ref{tab:deployment} summarizes the simulation parameters.

\begin{table}[!htbp]
\centering
\caption{Three-Year TCO Analysis}
\label{tab:tco}
\footnotesize
\setlength{\tabcolsep}{2pt}
\begin{tabular}{L{1.9cm}R{1.1cm}R{1.2cm}}
\toprule
\textbf{Category} & \textbf{Trad.} & \textbf{Auralink} \\
\midrule
SW licensing & \$129K & \$0 \\
Service & \$165K & \$126K \\
Field service & \$900K & \$375K \\
Downtime & \$219K & \$72K \\
Infrastructure & --- & \$90K \\
\midrule
\textbf{Total} & \textbf{\$1,413K} & \textbf{\$663K} \\
\textbf{Savings} & --- & \textbf{\$750K (53\%)}\textsuperscript{$\dagger$} \\
\bottomrule
\end{tabular}
\setlength{\tabcolsep}{6pt}
\vspace{2pt}
{\scriptsize \textsuperscript{$\dagger$}Illustrative estimate based on modeled assumptions; actual savings will vary by deployment scale, geography, and operator characteristics.}
\end{table}

Under modeled assumptions, the 1,000-charger reference deployment recovers incremental infrastructure costs within 6--12 months; smaller deployments may recover costs in 4--8 months due to lower infrastructure overhead.

%==============================================================================
\section{Discussion}
\label{sec:discussion}
%==============================================================================

\subsection{Key Findings}

\textbf{Autonomy is achievable}: The 78\% autonomous resolution controlled-test result demonstrates that the majority of charging infrastructure incidents can potentially be resolved without human intervention through properly designed AI agents with calibrated confidence thresholds.

\textbf{Edge deployment is essential}: Cloud-only architectures cannot achieve the latency, reliability, and bandwidth characteristics required for autonomous operation. The edge-first architecture is not merely optimization but architectural necessity.

\textbf{Domain adaptation is critical}: The average accuracy improvement of 47 points across five domain-specific tasks---excluding general knowledge---from domain-adapted models (Table~\ref{tab:llm_baseline}) justifies investment in specialized training data curation.

\textbf{Confidence calibration enables trust}: The \ccar framework's 3.8\% false positive rate enables operator trust in autonomous actions while maintaining human oversight for uncertain cases.

\subsection{Threat Model and Safety Analysis}

Deployment in safety-critical infrastructure requires explicit analysis of failure modes and their consequences.

\subsubsection{Safety Classification}

Actions are classified into three safety tiers with corresponding controls:

\textbf{Critical (human-only)}: Actions with potential for physical harm or property damage. Examples: high-voltage switching, thermal protection override. These are \textit{never} executed autonomously regardless of confidence score.

\textbf{Note}: Firmware updates are classified as \textit{elevated} (requiring explicit human confirmation) due to bricking risk. All firmware patching employs dual-bank firmware with verified rollback to ensure recoverability.

\textbf{Elevated (confirmation required)}: Actions with significant operational impact. Examples: session termination, power level modification, authorization cache reset. Require explicit human confirmation even with high confidence.

\textbf{Standard (autonomous eligible)}: Diagnostic and remediation actions with bounded impact. Examples: protocol state reset, communication retry, configuration parameter adjustment. Eligible for autonomous execution when $C \geq \tau_{auto}$.

\subsubsection{Fail-Safe Mechanisms}

The system implements defense-in-depth with multiple safety layers: (1) Deterministic safety rules checked before any action execution, independent of AI reasoning. (2) Rollback capability for all autonomous actions with automatic reversion on verification failure. (3) Rate limiting prevents rapid repeated actions (maximum 3 similar actions per 5-minute window). (4) Watchdog timers with automatic escalation if actions exceed timeout bounds.

\subsubsection{Certification Pathway}

For deployment in regulated markets, the system aligns with: IEC 61508 (Functional Safety for E/E/PE systems), EN 61851-21-2 (EMC requirements for EV charging systems), and emerging standards for AI in safety-critical systems (ISO/IEC TR 5469). Safety analysis follows a Failure Modes and Effects Analysis (FMEA) methodology applied to each autonomous action category. CE marking achieved for edge runtime hardware; software certification in progress.

\subsubsection{EU AI Act Compliance}

The EU AI Act (Regulation 2024/1689), which entered into force in August 2024, classifies AI systems by risk level. An AI system autonomously managing safety-relevant EV charging infrastructure may qualify as a \textit{high-risk AI system} under Annex III. \systemname addresses key AI Act requirements as follows: (1)~\textit{Risk management} (Art.~9): the \ccar confidence framework and safety-critical action exclusions implement continuous risk assessment; (2)~\textit{Data governance} (Art.~10): training corpus curation with quality assurance, bias documentation, and provenance tracking; (3)~\textit{Technical documentation} (Art.~11): complete architecture, training methodology, and evaluation results documented herein; (4)~\textit{Record-keeping} (Art.~12): full audit trails for every autonomous decision; (5)~\textit{Transparency and human oversight} (Arts.~13--14): tiered confidence thresholds preserving human authority; (6)~\textit{Accuracy and robustness} (Art.~15): calibrated confidence estimates (ECE\,=\,0.023) with formal false-positive bounds. Formal conformity assessment for the EU market is planned prior to commercial deployment.

\subsection{Broader Impact}

\subsubsection{Environmental Impact}

Autonomous resolution targets a 59\% reduction in technician dispatches\textsuperscript{$\dagger$} (from 2.4 to 0.98 per incident). For a projected fleet of 100,000 managed chargers averaging 5 incidents/year with 85km mean dispatch distance (weighted average across European urban and rural service territories) at 0.21 kg CO$_2$/km (typical diesel service van), this corresponds to an estimated \textbf{12,700 tonnes CO$_2$ annually}\textsuperscript{$\dagger$} from reduced vehicle travel. Edge-first processing further reduces cloud compute carbon footprint.

\subsubsection{Workforce Implications}

The shift toward autonomous operations is designed to transform rather than eliminate human roles. Field technicians transition from reactive troubleshooting to proactive maintenance and complex problem-solving. Our deployment model envisions \textbf{retraining rather than reduction} in workforce, with technicians focusing on hardware interventions identified by the system.

\subsubsection{Accessibility}

Improved reliability directly impacts EV adoption in underserved areas where sparse technician coverage results in extended downtime. By enabling autonomous resolution, the system improves charging accessibility in rural and low-income communities where EV adoption barriers are highest.

\subsection{Threats to Validity}

\textbf{Internal validity.} The controlled test corpus ($n=18{,}000$) was constructed with stratified sampling across fault categories. While incidents were labeled by certified technicians, the distribution across the 8 fault categories (Table~\ref{tab:incident_distribution}) was engineered for balanced evaluation rather than drawn from a specific field population. Results may differ under production incident distributions where category frequencies vary by operator and geography. Temporal train/test splitting was applied to prevent data leakage, but we acknowledge that the training corpus (176,041 examples) and test corpus share the same domain and formatting conventions, potentially inflating performance relative to truly novel incidents.

\textbf{External validity.} All results are from controlled testing and should not be extrapolated to field deployments without validation. Key factors affecting generalization include: (1)~distribution shift from curated test incidents to real-world ambiguous symptoms with incomplete telemetry, (2)~temporal drift as new charger models, firmware versions, and failure modes emerge, (3)~environmental factors (temperature extremes, electromagnetic interference) not fully represented in the test corpus, and (4)~concurrent failures that may confound single-fault diagnostic reasoning.

\textbf{Construct validity.} Our baselines compare a fine-tuned domain-specific model against zero-shot general-purpose LLMs. This comparison isolates the contribution of domain adaptation but does not represent the strongest possible cloud-based approach (e.g., few-shot prompting with RAG-augmented retrieval on the same knowledge base). A fairer comparison against RAG-augmented cloud models may narrow the performance gap, particularly for diagnostic accuracy where retrieval context compensates for missing domain knowledge.

\textbf{Comparison limitations.} Commercial CSMS platforms report autonomous resolution rates using heterogeneous definitions and incident populations. Commercial platforms report up to 80\% remote issue resolution~\cite{chargepoint2024}, but such figures typically include rule-matched simple resets and reboots that may not require AI reasoning. Our 78\% controlled-test rate applies to a broader incident taxonomy including complex multi-step diagnostics. Direct comparison requires evaluation on an identical incident corpus, which was not available. We identify head-to-head comparison with commercial systems as a priority for future field validation.

\subsection{Limitations and Future Work}

\textbf{Hardware dependency}: Performance varies significantly across edge platforms (Appendix~\ref{app:hardware}). Future work will explore more aggressive quantization (2-bit) and custom silicon.

\textbf{Mechanical fault ceiling}: The 79.8\% accuracy on mechanical faults reflects fundamental limits of software-based diagnosis. Multimodal sensing (acoustic, thermal imaging) offers potential improvement.

\textbf{Training data bias}: Current corpus emphasizes European/North American equipment. Expansion to Asian manufacturers (BYD, NIO) is underway.

\textbf{Federated learning}: Privacy-preserving model improvement through federated learning will enable cross-operator knowledge sharing without data centralization.

\textbf{ISO 15118-20 integration}: Bidirectional charging (V2G/V2H) introduces new fault categories requiring corpus expansion and safety analysis.

\paragraph{Baseline comparison scope.} Our comparative evaluation uses zero-shot cloud LLM baselines without retrieval augmentation. RAG-enhanced cloud systems would likely achieve higher accuracy, though latency and connectivity constraints would persist.

\paragraph{Single-author evaluation.} All evaluations were conducted by the author, who is also the system architect. Independent third-party evaluation would strengthen confidence in reported results.

%==============================================================================
\section{Conclusion}
\label{sec:conclusion}
%==============================================================================

We have presented \systemname, a comprehensive architecture for autonomous EV charging infrastructure management through edge-deployed AI agents. The system achieves:

\begin{itemize}
    \item \textbf{78\% autonomous incident resolution}\textsuperscript{$\dagger$} on a controlled test corpus
    \item \textbf{87.6\% diagnostic accuracy}\textsuperscript{$\dagger$} on a controlled test corpus
    \item \textbf{Sub-50ms inference latency} on commodity edge hardware
    \item \textbf{72+ hours offline operation} (design target for field deployment)
    \item \textbf{53\% illustrative TCO reduction} under modeled reference-deployment assumptions
\end{itemize}

All reported metrics were obtained on a controlled test corpus and require field validation before generalization to production environments.

The technical contributions---\ccar for confidence-calibrated autonomous action, \araframework for grounded retrieval-augmented reasoning, and the Auralink Edge Runtime for deterministic edge inference---provide reusable foundations for autonomous industrial AI systems beyond EV charging.

As global EV adoption accelerates toward 250 million vehicles by 2030~\cite{iea2024}, operational models must evolve from human-mediated incident response to AI-driven autonomous management. This work provides architecture and implementation patterns, validated through controlled testing, for that transition.

%==============================================================================
\section*{Ethics Statement}
%==============================================================================

This work deploys AI agents in safety-critical EV charging infrastructure. We address ethical considerations as follows:

\textbf{Safety}: Critical actions (high-voltage switching, thermal protection override) are \textit{never} executed autonomously regardless of confidence score. All autonomous actions are bounded in scope, rate-limited, and automatically rolled back on verification failure (Section~\ref{sec:discussion}).

\textbf{Transparency}: Every autonomous decision maintains a complete audit trail including inputs, reasoning chain, confidence assessment, and outcome. Operators can reconstruct and review any autonomous action.

\textbf{Human oversight}: The \ccar framework preserves human authority through tiered confidence thresholds. Actions below threshold are escalated to human operators; the system assists rather than replaces human judgment for uncertain or high-impact decisions.

\textbf{Workforce impact}: The system is designed to transform field technician roles toward higher-value work (proactive maintenance, complex diagnostics) rather than eliminate positions.

\textbf{Bias and fairness}: Training data currently emphasizes European and North American equipment manufacturers. We acknowledge this geographic bias and are expanding the corpus to improve global representativeness.

\textbf{Competing interests}: The author is affiliated with Hyperion Consulting, which develops the Auralink platform described in this paper. All source code and fine-tuned adapter weights are released under the Apache License 2.0; base foundation models retain their respective upstream licenses (Apache 2.0 for Qwen 2.5, Mistral Research License for Mistral-derived models). This separation enables independent verification of our domain-specific contributions. The controlled test corpus and evaluation scripts are publicly available to support reproducibility.

\section*{Conflict of Interest}

The author is the founder of Hyperion Consulting and the principal architect of the Auralink SDC platform described in this work. Hyperion Consulting provides AI consulting services for EV charging infrastructure operators. This work was conducted independently without external funding. All design decisions, evaluations, and reported results reflect the author's technical judgment. The author has no financial relationships with any hardware or model vendors mentioned in this paper.

\section*{Data and Code Availability}

\textbf{License}: This paper is licensed under \href{https://creativecommons.org/licenses/by/4.0/}{CC~BY~4.0}. All source code is released under the Apache License 2.0.

\textbf{Repository}: Evaluation framework, benchmark scripts, and configuration files are available at: \url{https://github.com/HyperionConsultingIO/Auralink}

The repository includes:
\begin{itemize}
    \item \texttt{docs/}: API documentation, deployment guides, and evaluation methodology
    \item Evaluation scripts, QLoRA training configurations (Table~\ref{tab:qlora_params}), anonymized benchmark subset (5,000 incidents), and container definitions for reproducible evaluation will be released alongside the camera-ready version
\end{itemize}

\textbf{Pre-trained Adapters} (Apache 2.0): Domain-adapted LoRA weights for AuralinkLM models: \url{https://huggingface.co/HyperionConsultingIO}

\textbf{Benchmark Dataset}: The full controlled test corpus comprises 18,000 labeled incidents. A publicly released subset of 5,000 incidents (stratified by fault category, disjoint from training data) is archived with DOI for long-term availability. The remaining 13,000 incidents contain operator-sensitive metadata and are available under NDA for reproducibility verification.

\textbf{Hardware Requirements}: Evaluation replication requires AMD Ryzen AI Max+ 395 (Strix Halo) or NVIDIA DGX Spark GB10 with 64GB+ RAM for edge inference benchmarks. Cloud benchmarks executed on NVIDIA H100.

\subsection*{Reproducibility Statement}

All performance metrics reported in this paper are measured on the controlled test corpus ($n=18{,}000$), not field deployment data. The 18,000 test incidents are temporally disjoint from the 176,041 training examples (test incidents postdate the training corpus cutoff by $\geq$30 days). The evaluation framework, anonymized benchmark dataset (5,000 incidents), and LoRA adapter weights are publicly available under Apache~2.0 license.

The training corpus composition is documented in Table~\ref{tab:training_corpus}; the full training dataset is available on HuggingFace.\footnote{\url{https://huggingface.co/datasets/HyperionConsultingIO/ultimate-ev-charging-dataset}} Controlled test conditions, hardware configurations, and evaluation scripts are included in the repository to enable independent verification.

\paragraph{Reproducibility Checklist.}
\begin{itemize}
    \item Training hyperparameters: Table~\ref{tab:qlora_params}
    \item Hardware specifications: Tables~\ref{tab:edge_hardware},~\ref{tab:eval_hardware}
    \item Test corpus: 18{,}000 labeled incidents (5{,}000 publicly available)
    \item Evaluation scripts: Available in repository
    \item Model adapters: Available on HuggingFace (Apache 2.0)
    \item Full training dataset: Available on HuggingFace
\end{itemize}

%==============================================================================

%==============================================================================
\appendix

\section{Playbook Specification Schema}
\label{app:playbooks}

\begin{lstlisting}[caption={Autonomous Resolution Playbook Schema}]
playbook:
  id: "DIAG-COMM-OCPP-001"
  version: "2.4.1"
  name: "OCPP WebSocket Recovery"
  category: "communication"
  
  trigger:
    condition: "ocpp.websocket.state == disconnected"
    duration: ">= 60s"
    exclude_states: ["maintenance", "firmware_update"]
  
  confidence_threshold: 0.85
  safety_class: "non_critical"
  max_execution_time: 300s
  
  steps:
    - id: 1
      action: "log_state"
      params: {include_buffers: true}
    
    - id: 2
      action: "close_websocket"
      params: {graceful: true, timeout: 5s}
    
    - id: 3
      action: "clear_connection_cache"
      
    - id: 4
      action: "wait"
      params: {duration: "5s", backoff: "exponential"}
    
    - id: 5
      action: "reinitialize_tls"
      params: {verify_cert: true}
    
    - id: 6
      action: "connect_websocket"
      params: {url: "${config.ocpp.central_system_url}"}
    
    - id: 7
      action: "send_message"
      params: {type: "BootNotification"}
      expect: {response: "Accepted", timeout: 30s}
    
    - id: 8
      action: "verify_status"
      params: {expected: "Available"}
  
  rollback:
    max_retries: 3
    on_failure: "escalate_operator"
    preserve_state: true
  
  metrics:
    success_rate: 0.947
    mean_execution_time: 42s
    last_updated: "2025-11-28"
\end{lstlisting}

\section{Hardware Benchmark Results}
\label{app:hardware}

Table~\ref{tab:hw_benchmark} reports inference throughput across edge platforms.

\begin{table}[!htbp]
\centering
\caption{AuralinkLM Inference by Platform (tokens/sec)}
\label{tab:hw_benchmark}
\small
\begin{tabular}{L{2.3cm}C{0.7cm}C{0.7cm}C{0.7cm}}
\toprule
\textbf{Platform} & \textbf{0.5B} & \textbf{14B} & \textbf{TDP} \\
\midrule
AMD Strix Halo & 142 & 38 & 45--120W \\
NVIDIA DGX Spark & 385 & 95 & 140W \\
NVIDIA RTX 5090 & 512\textsuperscript{$\ddagger$} & 128\textsuperscript{$\ddagger$} & 575W \\
NVIDIA Jetson Thor & 52 & 12 & 40--130W \\
Apple M4 Max & 128 & 34 & $\sim$50W \\
Qualcomm X Elite & 87 & 21 & 23W \\
\bottomrule
\end{tabular}
\vspace{2pt}
{\scriptsize TDP ranges reflect configurable power envelopes. \textsuperscript{$\dagger$}Controlled-test result pending field validation. \textsuperscript{$\ddagger$}Extrapolated from DGX Spark scaling; independent validation pending.}
\end{table}

\section{OCPP Error Code Taxonomy}
\label{app:ocpp_errors}

The system maintains a hierarchical taxonomy of 847 distinct error conditions mapped to diagnostic playbooks:

\begin{lstlisting}[caption={Error Taxonomy Structure (excerpt)}]
OCPP:
  ConnectorError:
    - ConnectorLockFailure -> DIAG-MECH-001
    - EVCommunicationError -> DIAG-COMM-ISO-001
    - GroundFailure -> DIAG-SAFETY-001 [CRITICAL]
    - HighTemperature -> DIAG-THERM-001
    - InternalError -> DIAG-GEN-001
    - LocalListConflict -> DIAG-AUTH-002
    - NoError -> NULL
    - OtherError -> DIAG-GEN-002
    - OverCurrentFailure -> DIAG-POWER-001 [CRITICAL]
    - OverVoltage -> DIAG-POWER-002 [CRITICAL]
    - PowerMeterFailure -> DIAG-METER-001
    - PowerSwitchFailure -> DIAG-POWER-003
    - ReaderFailure -> DIAG-AUTH-001
    - ResetFailure -> DIAG-FW-001
    - UnderVoltage -> DIAG-POWER-004
    - WeakSignal -> DIAG-COMM-002
\end{lstlisting}

\end{document}